\documentclass[aps,pra,amsmath,two column,amssymb,showpacs]{revtex4}
\usepackage{txfonts}
\usepackage{mathbbold}
\usepackage{epsfig,bm,dcolumn}
\usepackage{graphicx}
\usepackage{color}
\usepackage{overpic}

\begin{document}

\title{Two-band description of resonant superfluidity in atomic Fermi gases}

\author{Lianyi He,$^{1}$ Hui Hu,$^{2}$ Xia-Ji Liu$^{2}$}

\affiliation{1 Theoretical Division, Los Alamos National Laboratory, Los Alamos, New Mexico 87545, USA \\
2 Centre for Quantum and Optical Science, Swinburne University of Technology, Melbourne, Victoria, 3122, Australia}

\date{\today}

\begin{abstract}
Fermionic superfluidity in atomic Fermi gases across a Feshbach resonance is normally described by the atom-molecule theory, which treats the closed channel as a noninteracting point boson. In this work we present a theoretical description of the resonant superfluidity in analogy to the  two-band superconductors. We employ the underlying two-channel scattering model of Feshbach resonance where the closed channel is treated as a composite boson with binding energy $\varepsilon_0$ and the resonance is triggered by the microscopic interchannel coupling $U_{12}$.
The binding energy $\varepsilon_0$ naturally serves as an energy scale of the system, which has been sent to infinity in the atom-molecule theory. We show that the atom-molecule theory can be viewed as a leading-order low-energy effective theory of the underlying fermionic theory in the limit $\varepsilon_0\rightarrow\infty$ and $U_{12}\rightarrow0$, while keeping the phenomenological atom-molecule coupling finite. The resulting two-band description of the superfluid state is in analogy to the BCS theory of two-band superconductors. In the dilute limit $\varepsilon_0\rightarrow\infty$, the two-band description recovers precisely the atom-molecule theory. The two-band theory provides a natural approach to study the corrections because of a finite binding energy $\varepsilon_0$ in realistic experimental systems. For broad and moderate resonances, the correction is not important for current experimental densities. However, for extremely narrow resonance, we find that the correction becomes significant. The finite binding energy correction could be important for the stability of homogeneous polarized superfluid against phase separation in imbalanced Fermi gases across a narrow Feshbach resonance.
\end{abstract}


\maketitle

\section {Introduction}

It is widely accepted that a crossover from the BCS superfluidity to the Bose-Einstein condensation (BEC) of molecules can be realized in an attractive Fermi gas by tuning the attraction from weak to strong \cite{BCS-BEC}. This interesting phenomenon has been experimentally observed in ultracold Fermi gases of alkali-metal atoms \cite{EXP} (such as $^6$Li and $^{40}$K). In these experiments, the  attractive strength is effectively tuned by means of the Feshbach resonances (FRs). The basic mechanism of the FR is the coupling between different scattering channels in alkali-metal atoms in a magnetic field \cite{Feshbach,book}.

\begin{figure}[!htb]
\begin{center}
\includegraphics[width=8.4cm]{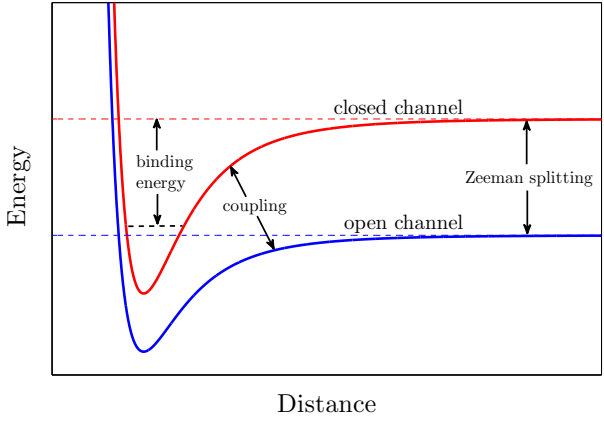}
\caption{(Color online) A schematic plot for the mechanism of FR. The red and blue solid lines show the potential energy (in proper units) as a function of the distance (in proper units) for the closed and open channels, respectively. The red and blue dashed lines show the scattering thresholds for the closed and open channels, respectively. The closed channel has a bound state with binding energy $\varepsilon_0$. This bound-state level can be tuned by changing the magnetic field. When it coincides with the scattering threshold of the open channel, a FR occurs. \label{fig1}}
\end{center}
\end{figure}

The scattering channels of the alkali-metal atoms are characterized by the eigenstates of the single-particle hyperfine Hamiltonian in a magnetic field $B$. The main contribution to the atom-atom interaction is the electrostatic central potential which also induces the couplings among different scattering channels. Because of these interchannel coupling, a FR occurs when the bound-state level of a certain closed channel coincides with the threshold of a certain open channel. A schematic plot for this mechanism is shown in Fig. \ref{fig1}. In the vicinity of an $s$-wave FR, the low-energy scattering amplitude for the open channel is given by
\begin{equation}
f(p)=\frac{1}{p\cot\delta(p)-ip},
\end{equation}
where the scattering phase shift can be well parametrized as \cite{Feshbach,book}
\begin{eqnarray}
p\cot\delta(p)=-\frac{1}{a_{\rm bg}}\frac{E-\gamma(B-B_0)}{E-\gamma(B-B_0)+\gamma B_\Delta}. \label{ScatteringAMP}
\end{eqnarray}
Here $E=p^2/M$ is the scattering energy, with $M$ being the atom mass, $B_0$ is the resonance point, $B_\Delta$ is the resonance width, $\gamma$ is the difference of the magnetic moments between the two-channels, and $a_{\rm bg}$ is the background scattering length. The units $\hbar=k_{\rm B}=1$ will be used throughout. The magnetic detuning $\delta(B)=\gamma(B-B_0)$ then tunes the effective scattering length of the open channel. Near the FR, $p\cot\delta(p)$ can be expanded as
\begin{equation}
p\cot\delta(p)\simeq-\frac{1}{a_{\rm eff}}+\frac{1}{2}r_{\rm eff}p^2+\cdots,
\end{equation}
with an effective scattering length
\begin{equation}
a_{\rm eff}=a_{\rm bg}\left(1-\frac{B_\Delta}{B-B_0}\right)
\end{equation}
and a negative effective range
\begin{equation}
r_{\rm eff}=-\frac{2}{Ma_{\rm bg}\gamma B_\Delta}.
\end{equation}
For many-body systems with total density $n$, we normally define a Fermi wave vector $k_{\rm F}=(3\pi^2n)^{1/3}$ and corresponding Fermi energy $\varepsilon_{\rm F}=k_{\rm F}^2/(2M)$. So far, most of the experimental studies focus on broad resonances, where $k_{\rm F}|r_{\rm eff}|\ll 1$. In this case, the many-body physics near the FR is universal and can be well described by a single-channel model. The universal many-body physics can be obtained from Monte Carlo simulations \cite{QMC} by using any short-ranged potential with the same scattering length $a_{\rm eff}$ and negligible effective range. Recently, resonantly interacting Fermi gas with a large effective range has been experimentally realized by using the narrow resonance of $^6$Li at $B\simeq543.3$G \cite{Expnarrow}.

For general resonances, a popular effective model is the atom-molecule model \cite{atom-molecule01,atom-molecule02,atom-molecule03,atom-molecule04}, which precisely reproduces the low-energy scattering amplitude parametrized by (\ref{ScatteringAMP}). The model Hamiltonian can be written as
\begin{equation}
H=H_{\rm f}+H_{\rm b}+H_{\rm fb},
\end{equation}
where the atom part,
\begin{equation}
H_{\rm f}=\sum_{\sigma=\uparrow,\downarrow}\int d^3{\bf r}\psi_{\sigma}^\dagger
\left(-\frac{\nabla^2}{2M}\right)\psi_{\sigma}^{\phantom{\dag}}+u_0\int d^3{\bf r}\psi_\uparrow^\dagger\psi_\downarrow^\dagger
\psi_\downarrow^{\phantom{\dag}}\psi_\uparrow^{\phantom{\dag}},
\end{equation}
the molecule part,
\begin{equation}
H_{\rm b}=\int d^3{\bf r}\phi_{\rm m}^\dagger \left(-\frac{\nabla^2}{4M}+\delta_0\right)\phi_{\rm m}^{\phantom{\dag}},
\end{equation}
and the atom-molecule coupling part,
\begin{equation}
H_{\rm bf}=g_0\int d^3{\bf r}\left(\phi_{\rm m}^\dagger\psi_\downarrow^{\phantom{\dag}}\psi_\uparrow^{\phantom{\dag}}
+\phi_{\rm m}^{\phantom{\dag}}\psi_\uparrow^\dagger\psi_\downarrow^\dagger\right).
\end{equation}
Here $\psi_{\sigma}$ denotes the open-channel fermions and $\phi_{\rm m}$ denotes the closed-channel molecules. The couplings $g_0$ and $u_0$ and the detuning $\delta_0$ are bare quantities. They should be renormalized by using the physical background scattering length $a_{\rm bg}$, resonance width $B_\Delta$, and detuning $\delta=\gamma(B-B_0)$. In this model, the closed channel
is treated as a point boson and the FR is triggered by the atom-molecule coupling $g_0$.

Another idea to study the narrow resonance is to use a well plus barrier potential \cite{barrier} which can reproduce a large and negative effective range. However, it is essentially a single-channel model which lacks the information of the closed channel. Actually, it has been shown that the closed channel dominates in the narrow resonance limit \cite{atom-molecule04}. In this paper, we go back to the underlying two-channel Hamiltonian which treats both the open and the closed channels as fermions \cite{Bruun}. We show that the (renormalized) atom-molecule coupling $g$ is related to the underlying inter-channel coupling $U_{12}$ and the closed-channel binding energy $\varepsilon_0$ through
\begin{eqnarray}
g=U_{12}\sqrt{\frac{\left(M\varepsilon_0\right)^{3/2}}{2\pi}}.
\end{eqnarray}
The binding energy $\varepsilon_0$ of the closed-channel bound state, which serves as a natural energy scale of the system, is automatically sent to infinity in the atom-molecule model. We show explicitly that the atom-molecule model can be viewed as a low-energy effective theory of the underlying two-channel theory in the limit $\varepsilon_0\rightarrow\infty$ and $U_{12}\rightarrow 0$, while keeping $g$ finite. For many-body physics, the resonant Fermi gas can be viewed as a two-band superfluid with a large band offset $\varepsilon_0$. Therefore, the underlying two-channel Hamiltonian will be referred to as a two-band model in this paper. In the dilute limit $\varepsilon_{\rm F}/\varepsilon_0\rightarrow0$, the prediction of the many-body physics becomes essentially the same as the atom-molecule model. However, in realistic experimental systems, the ratio $\varepsilon_{\rm F}/\varepsilon_0$ is small but finite. For broad and moderate resonances, the correction due to nonvanishing
$\varepsilon_{\rm F}/\varepsilon_0$ is not important. However, for extremely narrow resonance, this correction becomes significant.

The paper is organized as follows. In Sec. \ref{s2} we briefly review the atom-molecule model description of resonant superfluidity. In Sec. \ref{s3}
we calculate the low energy scattering amplitude in a two-band model and show that the atom-molecule model can be viewed as a low-energy effective theory. We formulate the resonant Fermi gas as a two-band superfluid in Sec. \ref{s4} and study its dilute limit in Sec. \ref{s5}. We apply the two-band description to study the narrow resonance of $^6$Li in Sec. \ref{s6}. The paper is summarized in Sec. \ref{s7}.

\section {Review: Atom-Molecule Theory}\label{s2}
In this section, we briefly review the atom-molecule theory of resonant superfluidity in atomic Fermi gases. We introduce the renormalization of the atom-molecule model and its description of the superfluid state \cite{atom-molecule02,atom-molecule03,atom-molecule04}.

\subsection {Renormalization of the model}

To renormalize the model, we first calculate the two-body scattering amplitude $f(p)$. The Lippmann-Schwinger equation for two-fermion scattering can be expressed by using an energy-dependent interaction vertex,
\begin{eqnarray}
V(E)=u_0+\frac{g_0^2}{E-\delta_0}.
\end{eqnarray}
The resulting $T$ matrix reads
\begin{eqnarray}
T(E)=\frac{V(E)}{1-V(E)\Pi(E)},
\end{eqnarray}
where the two-particle bubble function $\Pi(E)$ is given by
\begin{eqnarray}
\Pi(E)=\sum_{\bf k}\frac{1}{E+i\epsilon-2\varepsilon_{\bf k}}.
\end{eqnarray}
The integral over ${\bf k}$ is divergent and we introduce a cutoff $\Lambda$. Completing the integral we obtain
\begin{eqnarray}
\Pi(E)=-\frac{M\Lambda}{2\pi^2}+\frac{M}{4\pi}\sqrt{-M(E+i\epsilon)}.
\end{eqnarray}
The scattering amplitude $f(p)=-\frac{M}{4\pi}T(E)$ takes the form of Eq. (1), where $p\cot\delta(p)$ reads
\begin{eqnarray}
p\cot\delta(p)=-\frac{2\Lambda}{\pi}-\frac{4\pi}{M}\left(u_0+\frac{g_0^2}{E-\delta_0}\right)^{-1}.
\end{eqnarray}

Next we match the above result to the physical result (2). The renormalizability of the model requires that the equality
\begin{eqnarray}
&&-\frac{2\Lambda}{\pi}-\frac{4\pi}{M}\left(u_0(\Lambda)+\frac{g_0^2(\Lambda)}{E-\delta_0(\Lambda)}\right)^{-1}\nonumber\\
&=&-\frac{1}{a_{\rm bg}}\frac{E-\delta}{E-\delta+\gamma B_\Delta}
\end{eqnarray}
holds for an arbitrary value of the scattering energy $E$ through proper cutoff dependence of the bare couplings and the detuning. Defining the renormalized couplings $u=4\pi a_{\rm bg}/M$, $g=\sqrt{\gamma B_\Delta u}$, and detuning $\delta=\gamma(B-B_0)$, we obtain
\begin{eqnarray}
&&u_0(\Lambda)=\frac{u}{1-\eta(\Lambda)u},\nonumber\\
&&g_0(\Lambda)=\frac{g}{1-\eta(\Lambda)u},\nonumber\\
&&\delta_0(\Lambda)=\delta+\frac{g^2\eta(\Lambda)}{1-\eta(\Lambda)u},
\end{eqnarray}
where $\eta(\Lambda)=M\Lambda/(2\pi^2)$. When the background scattering length is neglected, i.e., $u=0$, only the detuning needs renormalization. In this case, we have $g_0=g$ and $\delta_0=\delta+g^2\eta(\Lambda)$.

\subsection {Superfluid state}

The partition function of the many-body system can be expressed as
\begin{eqnarray}
{\cal Z}=\int[d\psi][d\psi^\dagger][d\phi_{\rm m}^{\phantom{\dag}}][d\phi_{\rm m}^\dagger]\exp{\left(-{\cal S}_{\psi,\phi}\right)},
\end{eqnarray}
where the action ${\cal S}_{\rm eff}$ reads
\begin{eqnarray}
{\cal S}_{\psi,\phi}&=&\int dx\sum_{\sigma=\uparrow,\downarrow}\psi_{\sigma}^\dagger(x)
\left(\partial_\tau-\mu\right)\psi_{\sigma}^{\phantom{\dag}}(x)\nonumber\\
&+&\int dx \  \phi_{\rm m}^\dagger(x)\left(\partial_\tau-2\mu\right)\phi_{\rm m}^{\phantom{\dag}}(x)
+\int_0^\beta d\tau H.
\end{eqnarray}
Here $x=(\tau, {\bf r})$, with $\tau$ being the imaginary time, and $\beta=1/T$, with $T$ being the temperature of the system.  Here we have
introduced the chemical potential $\mu$ which is conjugate to the total particle number.
To decouple the four-fermion interaction term, we introduce an auxiliary field $\varphi(x)=u_0\psi_\downarrow(x)\psi_\uparrow(x)$.
By performing the Hubbard-Stratonovich transformation, we obtain
\begin{eqnarray}
{\cal Z}=\int[d\varphi][d\varphi^\dagger][d\phi_{\rm m}^{\phantom{\dag}}][d\phi_{\rm m}^\dagger]\exp{\left(-{\cal S}_{\rm eff}\right)}
\end{eqnarray}
where the effective action reads
\begin{eqnarray}
{\cal S}_{\rm eff}&=&\int dx \  \phi_{\rm m}^\dagger(x)\left(\partial_\tau-\frac{\nabla^2}{4M}+\delta_0-2\mu\right)
\phi_{\rm m}^{\phantom{\dag}}(x)\nonumber\\
&-&\int dx \ \frac{|\varphi(x)|^2}{u_0}-\int dx \ {\rm Tr}\ln {\bf G}^{-1}[\varphi,\phi_{\rm m}],
\end{eqnarray}
with the inverse fermion Green's function given by
\begin{eqnarray}
{\bf G}^{-1}=\left(\begin{array}{cc} -\partial_\tau+\frac{\nabla^2}{2M}+\mu& \varphi+g_0\phi_{\rm m}\\
\varphi^\dagger+g_0\phi_{\rm m}^\dagger& -\partial_\tau-\frac{\nabla^2}{2M}-\mu \end{array}\right)\delta(x-x^\prime).
\end{eqnarray}

In the superfluid phase, the two boson fields $\phi_{\rm m}$ and $\varphi$ generate nonzero expectation values. We define
\begin{eqnarray}
\Delta_{\rm b}=g_0\langle\phi_{\rm m}(x)\rangle,\ \ \ \ \ \ \Delta_{\rm f}=\langle\varphi(x)\rangle.
\end{eqnarray}
In the mean-field approximation, the grand potential at $T=0$ is given by
\begin{eqnarray}
\Omega_0=\frac{\delta_0-2\mu}{g_0^2}|\Delta_{\rm b}|^2-\frac{|\Delta_{\rm f}|^2}{u_0}
+\sum_{|{\bf k}|<\Lambda}\left(\xi_{\bf k}-E_{\bf k}\right),
\end{eqnarray}
where $\xi_{\bf k}=\varepsilon_{\bf k}-\mu$ and $E_{\bf k}=\sqrt{\xi_{\bf k}^2+|\Delta|^2}$, with $\Delta=\Delta_{\rm b}+\Delta_{\rm f}$. The next step is to remove the cutoff dependence by using the physical quantities $u$, $g$, and $\delta$.

To renormalize the grand potential, we note that $\Delta_{\rm b}$ and $\Delta_{\rm f}$ are cutoff dependent and therefore not physical quantities \cite{atom-molecule02}. To show this, we use the stationary condition
\begin{eqnarray}
&&\frac{\partial\Omega_0}{\partial\Delta_{\rm b}^*}=\frac{\delta_0-2\mu}{g_0^2}\Delta_{\rm b}-\sum_{\bf k}\frac{\Delta}{2E_{\bf k}}=0,\nonumber\\ &&\frac{\partial\Omega_0}{\partial\Delta_{\rm f}^*}=-\frac{\Delta_{\rm f}}{u_0}-\sum_{\bf k}\frac{\Delta}{2E_{\bf k}}=0
\end{eqnarray}
to obtain
\begin{eqnarray}
\Delta_{\rm b}=\frac{g_0^2}{2\mu-\delta_0}\frac{\Delta_{\rm f}}{u_0}.
\end{eqnarray}
Then we have
\begin{eqnarray}
&&\Delta_{\rm b}=\frac{g_0^2/(2\mu-\delta_0)}{u_0+g_0^2/(2\mu-\delta_0)}\Delta,\nonumber\\
&&\Delta_{\rm f}=\frac{u_0}{u_0+g_0^2/(2\mu-\delta_0)}\Delta.
\end{eqnarray}
Therefore, $\Delta_{\rm b}$ and $\Delta_{\rm f}$ are cutoff dependent. To renormalize the grand potential, we should regard them as dependent quantities and express the grand potential in terms of the finite quantity $\Delta$. Finally, we obtain
\begin{eqnarray}
\Omega_0(\Delta)=-\frac{|\Delta|^2}{u_0+g_0^2/(2\mu-\delta_0)}
+\sum_{|{\bf k}|<\Lambda}\left(\xi_{\bf k}-E_{\bf k}\right).
\end{eqnarray}
Using the fact that
\begin{eqnarray}
\frac{1}{u_0+g_0^2/(2\mu-\delta_0)}=\frac{1}{u_{\rm eff}}-\eta(\Lambda),
\end{eqnarray}
we obtain a cutoff-independent expression,
\begin{eqnarray}
\Omega_0(\Delta)=-\frac{|\Delta|^2}{u_{\rm eff}}+\sum_{\bf k}\left(\xi_{\bf k}-E_{\bf k}
+\frac{|\Delta|^2}{2\varepsilon_{\bf k}}\right),
\end{eqnarray}
where
\begin{eqnarray}
u_{\rm eff}=u+\frac{g^2}{2\mu-\delta}.
\end{eqnarray}

The gap equation can be derived from $\partial\Omega_0/\partial\Delta=0$. We have
\begin{eqnarray}
\frac{1}{u_{\rm eff}}=\sum_{\bf k}\left(\frac{1}{2\varepsilon_{\bf k}}-\frac{1}{2E_{\bf k}}\right).
\end{eqnarray}
Meanwhile, the total density $n$ is obtained through $n=-\partial\Omega_0/\partial\mu$. We obtain
\begin{eqnarray}
n=\sum_{\bf k}\left(1-\frac{\xi_{\bf k}}{E_{\bf k}}\right)+n_{\rm m},
\end{eqnarray}
where the contribution from the closed channel is given by
\begin{eqnarray}
n_{\rm m}=\frac{2|\Delta|^2}{g^2}\left(1-\frac{u}{u_{\rm eff}}\right)^2.
\end{eqnarray}
From the above coupled equations, we can solve the pairing gap $\Delta$, the chemical potential $\mu$, and the closed-channel
fraction $n_{\rm m}/n$ at given detuning $\delta$. For sufficiently large coupling $g$, the result reproduces the universality
predicted by the single-channel model. For finite temperature properties and beyond-mean-field treatment, we refer to
Refs. \cite{atom-molecule02,atom-molecule03,atom-molecule04}.

\section {A Two-Band Model for Feshbach Resonance}\label{s3}

The precise prediction of the FRs relies on solving the microscopic multichannel scattering problem with known microscopic interaction potentials. However, the scattering problem near a specific FR can be attributed to an effective two-channel problem. Let us consider a two-channel Hamiltonian $H=H_0+H_{\rm int}$ \cite{Bruun}, where
\begin{eqnarray}
H_0=\sum_{{\rm n}=1,2}\sum_{\sigma=\uparrow,\downarrow}\int d^3{\bf r}\psi_{{\rm n}\sigma}^\dagger({\bf r})
\left(-\frac{\nabla^2}{2M}+\varepsilon_{{\rm n}\sigma}\right)\psi_{{\rm n}\sigma}^{\phantom{\dag}}({\bf r}).
\end{eqnarray}
Here ${\rm n}=1$ and ${\rm n}=2$ correspond to the open channel and the closed channel, respectively. The interaction part is
\begin{eqnarray}
H_{\rm int}=\sum_{{\rm m},{\rm n}=1,2}\int d^3{\bf r} \int d^3{\bf r}^\prime\varphi_{\rm m}^\dagger({\bf r})
V_{{\rm mn}}(|{\bf r}-{\bf r}^\prime|)\varphi_{\rm n}^{\phantom{\dag}}({\bf r}^\prime),
\end{eqnarray}
where we use the notation
\begin{equation}
\varphi_{\rm n}({\bf r})=\psi_{{\rm n}\downarrow}({\bf r})\psi_{{\rm n}\uparrow}({\bf r}).
\end{equation}
In this second quantization form, the thresholds $\varepsilon_{\rm th}^{\rm n}=\varepsilon_{{\rm n}\uparrow}+\varepsilon_{{\rm n}\downarrow}$ are put into the free part $H_0$. Therefore, the interaction potential $V(|{\bf r}-{\bf r}^\prime|)\rightarrow0$ for $|{\bf r}-{\bf r}^\prime|\rightarrow\infty$. It includes both intra- and inter-channel interactions.

The threshold energies $\varepsilon_{{\rm n}\sigma}$ can be further simplified. Without loss of generality, we set
\begin{eqnarray}
\varepsilon_{1\uparrow}=\varepsilon_{1\downarrow}=0,\ \ \ \ \ \ \varepsilon_{2\uparrow}=\varepsilon_{2\downarrow}=\frac{1}{2}\varepsilon_{\rm th}.
\end{eqnarray}
For a many-body system, the difference between $\varepsilon_{{\rm n}\uparrow}$ and $\varepsilon_{{\rm n}\downarrow}$ can be absorbed into the definition of the chemical potentials.

\subsection {Low-energy scattering amplitude}

The effective range $r_0$ of the microscopic potential $V(|{\bf r}-{\bf r}^\prime|)$ introduces an energy scale
\begin{equation}
\varepsilon_{\rm r}=\frac{1}{Mr_0^2}.
\end{equation}
At low scattering energy $E\ll\varepsilon_{\rm r}$, the shape of the microscopic interaction potential $V(|{\bf r}-{\bf r}^\prime|)$ is not important. It can be safely replaced with a contact one $V\delta({\bf r}-{\bf r}^\prime)$. For many-body physics, this means that all kinds of short-ranged potential $V(|{\bf r}-{\bf r}^\prime|)$ leads to the same predictions in the dilute limit $k_{\rm F}r_0\rightarrow0$ \cite{Bruun}.
By making use of the contact potential, the Lippmann-Schwinger equation of the scattering $T$ matrix becomes an algebra equation,
\begin{eqnarray}
\left(\begin{array}{cc} T_{11}(E)& T_{12}(E)\\ T_{21}(E)& T_{22}(E)\end{array}\right)^{-1}=
\left(\begin{array}{cc} V_{11}& V_{12}\\ V_{21}& V_{22}\end{array}\right)^{-1}-
\left(\begin{array}{cc} {\cal B}_1(E)& 0\\ 0& {\cal B}_2(E)\end{array}\right), \label{Lippmann}
\end{eqnarray}
where the two-particle bubble functions are given by
\begin{equation}
{\cal B}_{\rm n}(E)=\sum_{\bf k}\frac{1}{E+i\epsilon-\varepsilon_{\rm th}^{\rm n}-2\varepsilon_{\bf k}}.
\end{equation}
Here $\epsilon=0^+$ and $\varepsilon_{\bf k}={\bf k}^2/(2M)$. Note that we have set $\varepsilon_{\rm th}^1=0$ and $\varepsilon_{\rm th}^2\equiv \varepsilon_{\rm th}(B)$ without loss of generality. The cost of the contact interaction is that the integral over the fermion momentum ${\bf k}$  becomes divergent. We introduce a cutoff $\Lambda$ for $|{\bf k}|$ and obtain
\begin{equation}
{\cal B}_{\rm n}(E)=-\frac{M\Lambda}{2\pi^2}+\Pi_{\rm n}(E),
\end{equation}
where
\begin{eqnarray}
&&\Pi_1(E)=\frac{M}{4\pi}\sqrt{-M(E+i\epsilon)},\nonumber\\
&&\Pi_2(E)=\frac{M}{4\pi}\sqrt{-M(E+i\epsilon-\varepsilon_{\rm th})}.
\end{eqnarray}
The divergence can be removed by using the renormalized coupling matrix $U$. It is related to the bare coupling matrix $V$ by \cite{Bruun}
\begin{eqnarray}
\left(\begin{array}{cc} U_{11}& U_{12}\\ U_{21}& U_{22}\end{array}\right)^{-1}=
\left(\begin{array}{cc} V_{11}& V_{12}\\ V_{21}& V_{22}\end{array}\right)^{-1}+
\left(\begin{array}{cc} \eta(\Lambda)& 0\\ 0& \eta(\Lambda)\end{array}\right). \label{Lippmann}
\end{eqnarray}
Without loss of generality, we set $U_{12}=U_{21}>0$. Then the Lippmann-Schwinger equation becomes cutoff independent,
\begin{eqnarray}
\left(\begin{array}{cc} T_{11}(E)& T_{12}(E)\\ T_{21}(E)& T_{22}(E)\end{array}\right)^{-1}=
\left(\begin{array}{cc} U_{11}& U_{12}\\ U_{21}& U_{22}\end{array}\right)^{-1}-
\left(\begin{array}{cc} \Pi_1(E)& 0\\ 0& \Pi_2(E)\end{array}\right).
\end{eqnarray}

Next we relate the elements of $U$ to physical observables. In general, both the coupling $U$ and the threshold energy $\varepsilon_{\rm th}$ depend on the magnetic field $B$. However, near the FR we may safely neglect the $B$ dependence of the coupling $U$. The threshold energy $\varepsilon_{\rm th}$ can be well parametrized as
\begin{equation}
\varepsilon_{\rm th}(B)=\varepsilon_0+\delta(B),
\end{equation}
where $\varepsilon_0$ is the binding energy of the closed-channel molecule and $\delta(B)=\gamma(B-B_0)$ is the magnetic detuning. The binding energy $\varepsilon_0$ serves as another energy scale of the system. For an atomic system, we normally have the hierarchy $\varepsilon_0\ll\varepsilon_{\rm r}$. For the problem of FR, low-energy scattering means that the scattering energy $E\ll\varepsilon_0$. This is actually the simplest model for FR in atomic systems. If we know the explicit $B$ dependence of the microscopic interaction potential $V(|{\bf r}-{\bf r}^\prime|)$ and the threshold energy $\varepsilon_{\rm th}$, we can have better description of the $B$ dependence \cite{Bruun}.

Solving the Lippmann-Schwinger equation, we obtain the $T$ matrix for the open channel,
\begin{eqnarray}
T_{11}^{-1}(E)=\left[U_{11}+\frac{U_{12}^2\Pi_2(E)}{1-U_{22}\Pi_2(E)}\right]^{-1}-\Pi_1(E).
\end{eqnarray}
A FR occurring at $B=B_0$ requires that $T_{11}(E=0)$ diverges at $B=B_0$. Since $\Pi_1(0)=0$, we obtain
\begin{eqnarray}
\frac{1}{U_{22}}=\Pi_2(0)=\frac{M}{4\pi}\sqrt{M\varepsilon_0}.
\end{eqnarray}
This equation clearly shows that the bound-state level of the closed channel coincides with the threshold of the open channel when FR occurs.
The scattering amplitude for the open channel is defined as $f(p)= -\frac{M}{4\pi}T_{11}(E)$. At low scattering energy
$E=p^2/M\ll \varepsilon_0$, $\Pi_2(E)$ is real and $\Pi_1(E)=-\frac{M}{4\pi}ip$. Therefore, $f(p)$ takes the form of Eq. (1),
where $p\cot\delta(p)$ is given by
\begin{widetext}
\begin{eqnarray}
p\cot\delta(p)
=-\frac{4\pi}{MU_{11}}\frac{\sqrt{M\varepsilon_0}-\sqrt{M(\varepsilon_0+\delta-E)}}
{\sqrt{M\varepsilon_0}-\sqrt{M(\varepsilon_0+\delta-E)}+\frac{U_{12}^2}{U_{11}U_{22}}\sqrt{M(\varepsilon_0+\delta-E)}}.
\end{eqnarray}
\end{widetext}
At low scattering energy $E\ll\varepsilon_0$ and near the FR ($\delta\ll\varepsilon_0$), it can be well approximated as
\begin{eqnarray}
p\cot\delta(p)\simeq-\frac{4\pi}{MU_{11}}\frac{E-\delta}
{E-\delta+\frac{2U_{12}^2}{U_{11}U_{22}}\varepsilon_0}.
\end{eqnarray}
Thus, the coupling constants are related to the physical observables through the following relations:
\begin{eqnarray}
U_{11}=\frac{4\pi a_{\rm bg}}{M},\ \ \ U_{22}=\frac{4\pi}{M}\frac{1}{\sqrt{M\varepsilon_0}}, \ \ \
\gamma B_\Delta=\frac{2U_{12}^2}{U_{11}U_{22}}\varepsilon_0.
\end{eqnarray}
In terms of $U_{12}$ and $\varepsilon_0$, the effective range $r_{\rm eff}$ can be explicitly expressed as
\begin{eqnarray}
r_{\rm eff}=-\frac{16\pi^2}{M^2U_{12}^2(M\varepsilon_0)^{3/2}},
\end{eqnarray}
which indicates that the effective range is always negative. From Fig. \ref{fig1}, we find that the binding energy $\varepsilon_0$ equals the
Zeeman energy splitting $E_{\rm Z}$ at the resonance $B=B_0$ \cite{Bruun}; i.e.,
\begin{eqnarray}
\varepsilon_0=E_{\rm Z}(B=B_0).
\end{eqnarray}

\subsection {Atom-molecule model as a low-energy effective theory}

The phenomenological coupling $g$ in the atom-molecule model is related to physical observables as $g=\sqrt{\gamma B_\Delta u}$. From Eq. (51) we can identify $u=U_{11}$. Therefore, $g$ can be expressed in terms of $\varepsilon_0$ and $U_{12}$ as
\begin{equation}
g=U_{12}\sqrt{\frac{(M\varepsilon_0)^{3/2}}{2\pi}}.
\end{equation}
This expressions shows explicitly how the phenomenological coupling $g$ is related to the microscopic parameters. In the following we show that the atom-molecule model can be viewed as a low-energy effective theory in the limit $\varepsilon_0\rightarrow\infty$ while keeping the phenomenological coupling $g$ finite (hence, $U_{12}\rightarrow0$). In this limit, we have $U_{12}\sim O(\varepsilon_0^{-3/4})$,
$U_{22}\sim O(\varepsilon_0^{-1/2})$, and $U_{11}\sim O(1)$,  which leads to $U_{12}^2\ll |U_{11}U_{22}|$. Therefore, the relations between $U$ and $V$ can be well approximated as
\begin{eqnarray}
\left(V_{11}-\frac{V_{12}^2}{V_{22}}\right)^{-1}&=&\left(U_{11}-\frac{U_{12}^2}{U_{22}}\right)^{-1}-\eta(\Lambda)\nonumber\\
&\simeq&\frac{1}{U_{11}}-\eta(\Lambda),\nonumber\\
\left(V_{22}-\frac{V_{12}^2}{V_{11}}\right)^{-1}&=&\left(U_{22}-\frac{U_{12}^2}{U_{11}}\right)^{-1}-\eta(\Lambda)\nonumber\\
&\simeq&\frac{1}{U_{22}}\left(1+\frac{U_{12}^2}{U_{11}U_{22}}\right)-\eta(\Lambda),\nonumber\\
\frac{V_{12}}{V_{22}}\left(V_{11}-\frac{V_{12}^2}{V_{22}}\right)^{-1}&=&\frac{U_{12}}{U_{11}U_{22}-U_{12}^2}
\simeq\frac{U_{12}}{U_{11}U_{22}}.
\end{eqnarray}
Comparing with the atom-molecule model, we identify
\begin{eqnarray}
u=U_{11},\ \ \ \ \ \ u_0=V_{11}-\frac{V_{12}^2}{V_{22}}.
\end{eqnarray}

To arrive at the atom-molecule model we introduce an auxiliary field
$\Phi_{\rm m}(x)=V_{22}\psi_{2\downarrow}(x)\psi_{2\uparrow}(x)$ and integrate out the closed-channel fermions.
Then the effective action can be expressed as ${\cal S}_{\rm eff}={\cal S}_{\rm f}+{\cal S}_{\rm b}+{\cal S}_{\rm bf}$, where
\begin{widetext}
\begin{eqnarray}
{\cal S}_{\rm f}&=&\int dx \left[\sum_{\sigma=\uparrow,\downarrow}\psi_{\sigma}^\dagger(x)
\left(\partial_\tau-\frac{\nabla^2}{2M}-\mu\right)\psi_{\sigma}^{\phantom{\dag}}(x)+\left(V_{11}-\frac{V_{12}^2}{V_{22}}\right)\psi_\uparrow^\dagger(x)\psi_\downarrow^\dagger(x)
\psi_\downarrow^{\phantom{\dag}}(x)\psi_\uparrow^{\phantom{\dag}}(x)\right],\nonumber\\
{\cal S}_{\rm b}&=&\int dx\left[-\frac{|\Phi_{\rm m}(x)|^2}{V_{22}}-{\rm Trln}\left(\begin{array}{cc} -\partial_\tau+\frac{\nabla^2}{2M}+\mu-\frac{1}{2}\varepsilon_{\rm th}& \Phi_{\rm m}(x)\\
\Phi_{\rm m}^\dagger(x) & -\partial_\tau-\frac{\nabla^2}{2M}-\mu+\frac{1}{2}\varepsilon_{\rm th} \end{array}\right)\right],\nonumber\\
{\cal S}_{\rm bf}&=&\int dx \frac{V_{12}}{V_{22}}\left[\Phi_{\rm m}(x)\psi_{\uparrow}^\dagger(x)\psi_{\downarrow}^\dagger(x)
+\Phi_{\rm m}^\dagger(x)\psi_{\downarrow}(x)\psi_{\uparrow}(x)\right].
\end{eqnarray}
\end{widetext}
Here we have introduced the chemical potential $\mu$ and used $\psi_\sigma\equiv\psi_{1\sigma}$ to denote the open-channel fermions. Using the fact
$V_{11}-V_{12}^2/V_{22}=u_0$ we find that the fermion part ${\cal S}_{\rm f}$ corresponds precisely to the atom part $H_{\rm f}$ of the atom-molecule model.

Next we consider the molecule part ${\cal S}_{\rm b}$ and the atom-molecule coupling part ${\cal S}_{\rm bf}$.  The inverse propagator for the boson field $\Phi(x)$ is given by
\begin{equation}
D_{\rm m}^{-1}(x,x^\prime)=\frac{\delta^2{\cal S}_{\rm b}[\Phi_{\rm m}^\dagger,\Phi_{\rm m}]}
{\delta\Phi_{\rm m}^\dagger(x)\delta\Phi_{\rm m}(x^\prime)}.
\end{equation}
In the momentum space, it can be explicitly evaluated as
\begin{eqnarray}
D_{\rm m}^{-1}(\omega,{\bf q})=-\frac{1}{V_{22}}+\sum_{\bf k}\frac{1}{\omega+i\epsilon+2\mu-\frac{1}{2}\varepsilon_{\bf q}
-\varepsilon_{\rm th}-2\varepsilon_{\bf k}}.
\end{eqnarray}
At low energy, i.e., $\omega,\varepsilon_{\bf q}\ll \varepsilon_0$, it can be expanded in terms of $\omega$ and $\varepsilon_{\bf q}$. We have
\begin{eqnarray}
D_{\rm m}^{-1}(\omega,{\bf q})\simeq d_0+d_1\left(\omega-\frac{{\bf q}^2}{4M}\right),
\end{eqnarray}
where
\begin{eqnarray}
d_0&=&\frac{M}{4\pi}\sqrt{M(\varepsilon_{\rm th}-2\mu)}-\left(\frac{1}{V_{22}}+\frac{M\Lambda}{2\pi^2}\right),\nonumber\\
d_1&=&\frac{M^2}{8\pi\sqrt{M(\varepsilon_{\rm th}-2\mu)}}.
\end{eqnarray}
It becomes evident in the following that the low-energy expansion (59) corresponds to the leading-order expansion in $1/\sqrt{M\varepsilon_0}$.
For large $\varepsilon_0$, we have $\delta,\mu\ll\varepsilon_0$. Therefore, $d_0$ and $d_1$ can be well approximated as
\begin{eqnarray}
d_1\simeq\alpha=\frac{M^2}{8\pi\sqrt{M\varepsilon_0}}
\end{eqnarray}
and
\begin{eqnarray}
d_0&\simeq&\frac{M}{4\pi}\sqrt{M(\varepsilon_{\rm th}-2\mu)}-\frac{1}{U_{22}}\nonumber\\
&&-\frac{U_{12}^2}{U_{11}U_{22}^2}+\frac{V_{12}^2}{V_{22}(V_{11}V_{22}-V_{12}^2)}\nonumber\\
&=&\frac{M}{4\pi}\left[\sqrt{M(\varepsilon_0+\delta-2\mu)}-\sqrt{M\varepsilon_0}\right]\nonumber\\
&&+\left(\frac{V_{12}}{V_{22}}\right)^2\frac{1}{u_0}-\left(\frac{U_{12}}{U_{22}}\right)^2\frac{1}{u}\nonumber\\
&\simeq&\alpha\left[\delta-2\mu+\frac{1}{\alpha u_0}\left(\frac{V_{12}}{V_{22}}\right)^2
-\frac{1}{\alpha u}\left(\frac{U_{12}}{U_{22}}\right)^2\right].
\end{eqnarray}

Then we define a normalized molecule field,
\begin{eqnarray}
\phi_{\rm m}(x)=\sqrt{\alpha}\Phi_{\rm m}(x),
\end{eqnarray}
which corresponds to the molecule field used in the atom-molecule model. The effective actions become
\begin{eqnarray}
{\cal S}_{\rm b}&\simeq&\int dx\ \phi_{\rm m}^\dagger(x)\Bigg[\partial_\tau-\frac{\nabla^2}{4M}-2\mu\nonumber\\
&&\ \ \ \ \ \ \ \ \ \ +\delta+\frac{1}{\alpha u_0}\left(\frac{V_{12}}{V_{22}}\right)^2
-\frac{1}{\alpha u}\left(\frac{U_{12}}{U_{22}}\right)^2\Bigg]\phi_{\rm m}^{\phantom{\dag}}(x),\nonumber\\
{\cal S}_{\rm bf}&\simeq&\int dx \frac{1}{\sqrt{\alpha}}\frac{V_{12}}{V_{22}}
\left[\phi_{\rm m}^{\phantom{\dag}}(x)\psi_{\uparrow}^\dagger(x)\psi_{\downarrow}^\dagger(x)
+{\rm H. c.}\right].
\end{eqnarray}
Using the relations between $U$ and $V$ in Eq. (54) we obtain
\begin{eqnarray}
\frac{1}{\sqrt{\alpha}}\frac{V_{12}}{V_{22}}=\frac{1}{\sqrt{\alpha}}\frac{U_{12}}{U_{11}U_{22}}u_0=\frac{g}{1-\eta(\Lambda)u}=g_0
\end{eqnarray}
and
\begin{eqnarray}
&&\delta+\frac{1}{\alpha u_0}\left(\frac{V_{12}}{V_{22}}\right)^2-\frac{1}{\alpha u}\left(\frac{U_{12}}{U_{22}}\right)^2\nonumber\\
&=&\delta+\frac{g_0^2}{u_0}-\frac{g^2}{u}=\delta+\frac{g^2\eta(\Lambda)}{1-\eta(\Lambda)u}=\delta_0.
\end{eqnarray}
Here we have used the definition of the atom-molecule coupling
\begin{eqnarray}
g=\sqrt{\gamma B_\Delta u}=\sqrt{\frac{2U_{12}^2}{U_{22}}\varepsilon_0}=U_{12}\sqrt{\frac{(M\varepsilon_0)^{3/2}}{2\pi}}.
\end{eqnarray}

Therefore, we have shown that the atom-molecule model is a low-energy effective theory of the two-band model
in the limit $\varepsilon_0\rightarrow\infty$ (and hence $U_{12}\rightarrow0$), while keeping the phenomenological atom-molecule coupling
$g$ finite. In the atom-molecule model, the energy scale $\varepsilon_0$ is hidden and is automatically sent to infinity.

We can also work out the next-to-leading-order low-energy expansion of the molecule part ${\cal S}_{\rm b}$. It is quartic in $\phi_{\rm m}$ and corresponds to the two-body interaction of the closed-channel bound states. We have
\begin{eqnarray}
{\cal S}_{\rm b}^{\rm NLO}=\frac{1}{2}\frac{4\pi a_{\rm m}}{2M}\int dx \ |\phi_{\rm m}(x)|^4,
\end{eqnarray}
where $a_{\rm m}\simeq2/\sqrt{M\varepsilon_0}$ is the scattering length of the closed-channel molecules. In the limit $\varepsilon_0\rightarrow\infty$, this contribution can be safely neglected. However, for realistic systems, $\varepsilon_0$ is large but finite,
this term may be important for the stability of polarized superfluidity \cite{Zhai}.

\section {Resonant Fermi Gas as a Two-Band Superfluid}\label{s4}

Starting from the two-channel Hamiltonian (35) and (36), we naturally have a two-band description of the superfluid state which is analogous to the BCS theory of two-band superconductors \cite{two-band,two-band-iskin}. The molecule binding energy $\varepsilon_0$ appears explicitly in this theory as the band offset. In the dilute limit $\varepsilon_{\rm F}/\varepsilon_0\rightarrow0$, we expect that the two-band description recovers the atom-molecule model description.

\subsection {Superfluid Phase}

Following the standard field theoretical treatment, we introduce the auxiliary pairing fields
\begin{eqnarray}
\Phi(x)=\left(\begin{array}{c} \Phi_1(x)\\ \Phi_2(x) \end{array}\right)=
\left(\begin{array}{cc} V_{11}& V_{12}\\ V_{21}& V_{22}\end{array}\right)
\left(\begin{array}{c} \varphi_1(x)\\ \varphi_2(x) \end{array}\right),
\end{eqnarray}
where $x=(\tau,{\bf r})$, with $\tau$ being the imaginary time, apply the Hubbard-Stratonovich transformation, and integrate out the fermion fields. The partition function of the system can be expressed as
\begin{equation}
{\cal Z}=\int[d\Phi^\dagger][d\Phi]\exp{\left(-{\cal S}_{\rm eff}\right)}.
\end{equation}
The effective action ${\cal S}_{\rm eff}$ reads
\begin{eqnarray}
{\cal S}_{\rm eff}=-\int dx \ \Phi^\dagger(x) V^{-1}\Phi(x)-\sum_{{\rm n}=1,2}{\rm Tr}\ln {\bf G}_{\rm n}^{-1}[\Phi_{\rm n}(x)],
\end{eqnarray}
where the inverse fermion Green's functions are given by
\begin{eqnarray}
{\bf G}_{\rm n}^{-1}=\left(\begin{array}{cc} -\partial_\tau+\frac{\nabla^2}{2M}+\mu_{\rm n}& \Phi_{\rm n}(x)\\
\Phi_{\rm n}^*(x)& -\partial_\tau-\frac{\nabla^2}{2M}-\mu_{\rm n} \end{array}\right)\delta(x-x^\prime).
\end{eqnarray}
Here we have defined $\mu_1=\mu$ and $\mu_2=\mu-\varepsilon_{\rm th}/2$, with $\mu$ being the fermion chemical potential.

In the superfluid phase, the pairing fields have nonzero expectation values. We write
\begin{equation}
\Phi_{\rm n}(x)=\Delta_{\rm n}+\phi_{\rm n}(x),
\end{equation}
where the constants $\Delta_1$ and $\Delta_2$ serve as the order parameters of superfluidity. Note that both $\Phi_1$ and $\Phi_2$ are superpositions of the pair potentials $\varphi_1$ and $\varphi_2$. The order parameters $\Delta_1$ and $\Delta_2$ are both finite quantities, in contrast to the atom-molecule model. The effective action can be expanded in terms of the fluctuations $\phi_{\rm n}(x)$. In the following, we evaluate the effective action up to the Gaussian fluctuations, i.e., ${\cal S}_{\rm eff}\simeq{\cal S}_0+{\cal S}_{\rm g}$. First, we consider the mean-field part ${\cal S}_0$. It can be evaluated as ${\cal S}_0=\beta {\cal V}\Omega_0$, where the grand potential $\Omega_0$ is given by
\begin{eqnarray}
\Omega_0&=&-\Delta^\dagger U^{-1}\Delta+\sum_{{\rm n}=1,2}\sum_{\bf k}\left(\xi_{{\rm n}\bf k}-E_{{\rm n}\bf k}
+\frac{|\Delta_{\rm n}|^2}{2\varepsilon_{\bf k}}\right)\nonumber\\
&-&2T\sum_{{\rm n}=1,2}\sum_{\bf k}\ln{\left(1+e^{-E_{{\rm n}\bf k}/T}\right)}.
\end{eqnarray}
Here $\Delta=(\Delta_1,\Delta_2)^{\rm T}$ and the dispersions are defined as $\xi_{{\rm n}\bf k}=\varepsilon_{\bf k}-\mu_{\rm n}$ and
$E_{{\rm n}\bf k}=\sqrt{(\xi_{{\rm n}\bf k})^2+|\Delta_{\rm n}|^2}$. Note that we have used the renormalized coupling $U$. The grand potential $\Omega_0$ here is free from the cutoff $\Lambda$ for arbitrary values of $\Delta_1$ and $\Delta_2$. Therefore, $\Delta_1$ and $\Delta_2$ are two independent physical quantities in the present two-band theory.

The contribution from Gaussian fluctuations is given by
\begin{equation}
{\cal S}_{\rm g}=\frac{1}{2}\sum_Q\phi^\dagger(-Q){\bf M}(Q)\phi(Q),
\end{equation}
where $Q=(i\omega_\nu,{\bf q})$ with $\omega_\nu=2\nu\pi/T$ ($\nu$ integer) and $\phi^\dagger(-Q)=(\phi_1^*(Q),\phi_1(-Q),\phi_2^*(Q),\phi_2(-Q))$. The inverse boson propagator ${\bf M}(Q)$ is a $4\times4$ matrix and can be expressed as
\begin{equation}
{\bf M}(Q)=-U^{-1}\otimes I_2+{\bf H}(Q),
\end{equation}
where $I_2$ is a $2\times2$ identity matrix. The matrix ${\bf H}(Q)$ can be expressed as
\begin{equation}
{\bf H}(Q)={\rm diag}(H_1(Q),H_2(Q)).
\end{equation}
The two blocks $H_{\rm n}(Q)$ are $2\times 2$ matrices. Their elements satisfies $H_{\rm n}^{11}(Q)=H_{\rm n}^{22}(-Q)$ and
$H_{\rm n}^{12}(Q)=H_{\rm n}^{21}(Q)$. Using the fermion propagator ${\cal G}_{\rm n}(K)$, we have
\begin{eqnarray}
&&H_{\rm n}^{11}(Q)=\sum_{K}{\cal G}_{\rm n}^{22}(K){\cal G}_{\rm n}^{11}(K+Q),\nonumber\\
&&H_{\rm n}^{12}(Q)=\sum_{K}{\cal G}_{\rm n}^{12}(K){\cal G}_{\rm n}^{21}(K+Q),
\end{eqnarray}
where ${\cal G}_{\rm n}(K)$ can be obtained from
\begin{eqnarray}
{\cal G}_{\rm n}^{-1}(K)=\left(\begin{array}{cc} i\omega_m-\xi_{{\rm n}{\bf k}}& \Delta_{\rm n}\\
\Delta_{\rm n}^*& i\omega_m-\xi_{{\rm n}{\bf k}}\end{array}\right).
\end{eqnarray}
Here $K=(i\omega_m,{\bf k})$, with $\omega_m=(2m+1)\pi/T$ ($m$ integer).
Their explicit forms are given by
\begin{widetext}
\begin{eqnarray}
H_{{\rm n}11}(Q)&=&\sum_{\bf k}\Bigg[(1-f_{{\rm n}+}-f_{{\rm n}-})
\left(\frac{u_{{\rm n}+}^2u_{{\rm n}-}^2}{i\omega_\nu-E_{{\rm n}+}-E_{{\rm n}-}}
-\frac{\upsilon_{{\rm n}+}^2\upsilon_{{\rm n}-}^2}{i\omega_\nu+E_{{\rm n}+}+E_{{\rm n}-}}\right)+\frac{1}{2\varepsilon_{\bf k}}\nonumber\\
&&\ \ \ \ \ \ +(f_{{\rm n}+}-f_{{\rm n}-})\left(\frac{\upsilon_{{\rm n}+}^2u_{{\rm n}-}^2}{i\omega_\nu+E_{{\rm n}+}-E_{{\rm n}-}}
-\frac{u_{{\rm n}+}^2\upsilon_{{\rm n}-}^2}{i\omega_\nu-E_{{\rm n}+}+E_{{\rm n}-}}\right)\Bigg],\nonumber\\
H_{{\rm n}12}(Q)&=&\sum_{\bf k}\frac{|\Delta_{\rm n}|^2}{2E_{{\rm n}+}E_{{\rm n}-}}\left[(1-f_{{\rm n}+}-f_{{\rm n}-})
\frac{E_{{\rm n}+}+E_{{\rm n}-}}{(E_{{\rm n}+}+E_{{\rm n}-})^2-(i\omega_\nu)^2}
+(f_{{\rm n}+}-f_{{\rm n}-})
\frac{E_{{\rm n}+}-E_{{\rm n}-}}{(E_{{\rm n}+}-E_{{\rm n}-})^2-(i\omega_\nu)^2}\right].
\end{eqnarray}
\end{widetext}
The notations in the above expressions are defined as $E_{{\rm n}\pm}=E_{{\rm n}{\bf k}\pm{\bf q}/2}$,
$u_{{\rm n}\pm}^2=\frac{1}{2}\left(1+\xi_{{\rm n}\pm}/E_{{\rm n}\pm}\right)$,
$\upsilon_{{\rm n}\pm}^2=\frac{1}{2}\left(1-\xi_{{\rm n}\pm}/E_{{\rm n}\pm}\right)$, and $f_{{\rm n}\pm}=f(E_{{\rm n}\pm})$, with
$f(E)\equiv 1/(e^{E/T}+1)$ being the Fermi-Dirac distribution. The contribution of the Gaussian fluctuations to the grand potential
can be formally expressed as
\begin{equation}
\Omega_{\rm g}=\frac{1}{2\beta}\sum_Q\ln\det {\bf M}(Q).
\end{equation}

The order parameters $\Delta_{\rm n}$ and the chemical potential $\mu$ should be determined by the stationary condition or the gap equation $\partial\Omega_0/\partial\Delta_{\rm n}=0$ together with the constraint for the total density $n=-\partial\Omega_{\rm t}/\partial\mu$, where
$\Omega_{\rm t}=\Omega_0+\Omega_{\rm g}$ is the grand potential including Gaussian fluctuations \cite{Hu}. The gap equation can be
expressed as
\begin{eqnarray}
\left[\left(\begin{array}{cc} U_{11}& U_{12}\\ U_{21}& U_{22}\end{array}\right)^{-1}-
\left(\begin{array}{cc} F_1(\Delta_1)& 0\\ 0& F_2(\Delta_2)\end{array}\right)\right]\left(\begin{array}{cc} \Delta_1\\ \Delta_2\end{array}\right)=0,
\end{eqnarray}
where $F_{\rm n}$ is given by
\begin{eqnarray}
F_{\rm n}(\Delta_{\rm n})=\sum_{\bf k}\left[\frac{2f(E_{{\rm n}\bf k})-1}{2E_{{\rm n}\bf k}}+\frac{1}{2\varepsilon_{\bf k}}\right].
\end{eqnarray}
We conclude that $\Delta_1$ and $\Delta_2$ vanish at the same critical temperature, in analogy to the BCS theory of two-band superconductors \cite{two-band}. Meanwhile, the number equation is given by
\begin{equation}
n=n_1+n_2+n_{\rm g},
\end{equation}
where $n_{\rm g}=-\partial\Omega_{\rm g}/\partial\mu$ is the fluctuation contribution and
\begin{eqnarray}
n_{\rm n}=\sum_{\bf k}\left[1-\frac{\xi_{{\rm n}\bf k}}{E_{{\rm n}\bf k}}\left(1-2f(E_{{\rm n}\bf k})\right)\right].
\end{eqnarray}
Note that the gap equation can also be expressed as
\begin{eqnarray}
&&\left[U_{11}+\frac{U_{12}^2F_2(\Delta_2)}{1-U_{22}F_2(\Delta_2)}\right]^{-1}=F_1(\Delta_1),\nonumber\\
&&\frac{\Delta_2}{\Delta_1}=\frac{U_{12}}{U_{11}-F_2(\Delta_2)\det U}.
\end{eqnarray}
The first equation shows explicitly the resonance effect on the open channel. As we show below, these equations become essentially the same
as the atom-molecule model in the dilute limit $\varepsilon_{\rm F}/\varepsilon_0\rightarrow0$.

\subsection {Superfluid Transition Temperature}
The superfluid order parameters $\Delta_1$ and $\Delta_2$ vanish simultaneously at some critical temperature $T_c$. At a given chemical potential $\mu$, the critical temperature is determined by
\begin{eqnarray}
\det\left[\left(\begin{array}{cc} U_{11}& U_{12}\\ U_{21}& U_{22}\end{array}\right)^{-1}-
\left(\begin{array}{cc} F_1(0)& 0\\ 0& F_2(0)\end{array}\right)\right]=0.
\end{eqnarray}
After some manipulations, we obtain
\begin{eqnarray}
\left[U_{11}+\frac{U_{12}^2F_2(0)}{1-U_{22}F_2(0)}\right]^{-1}=F_1(0).
\end{eqnarray}

\begin{figure*}[!htb]
\begin{center}
\includegraphics[width=8cm]{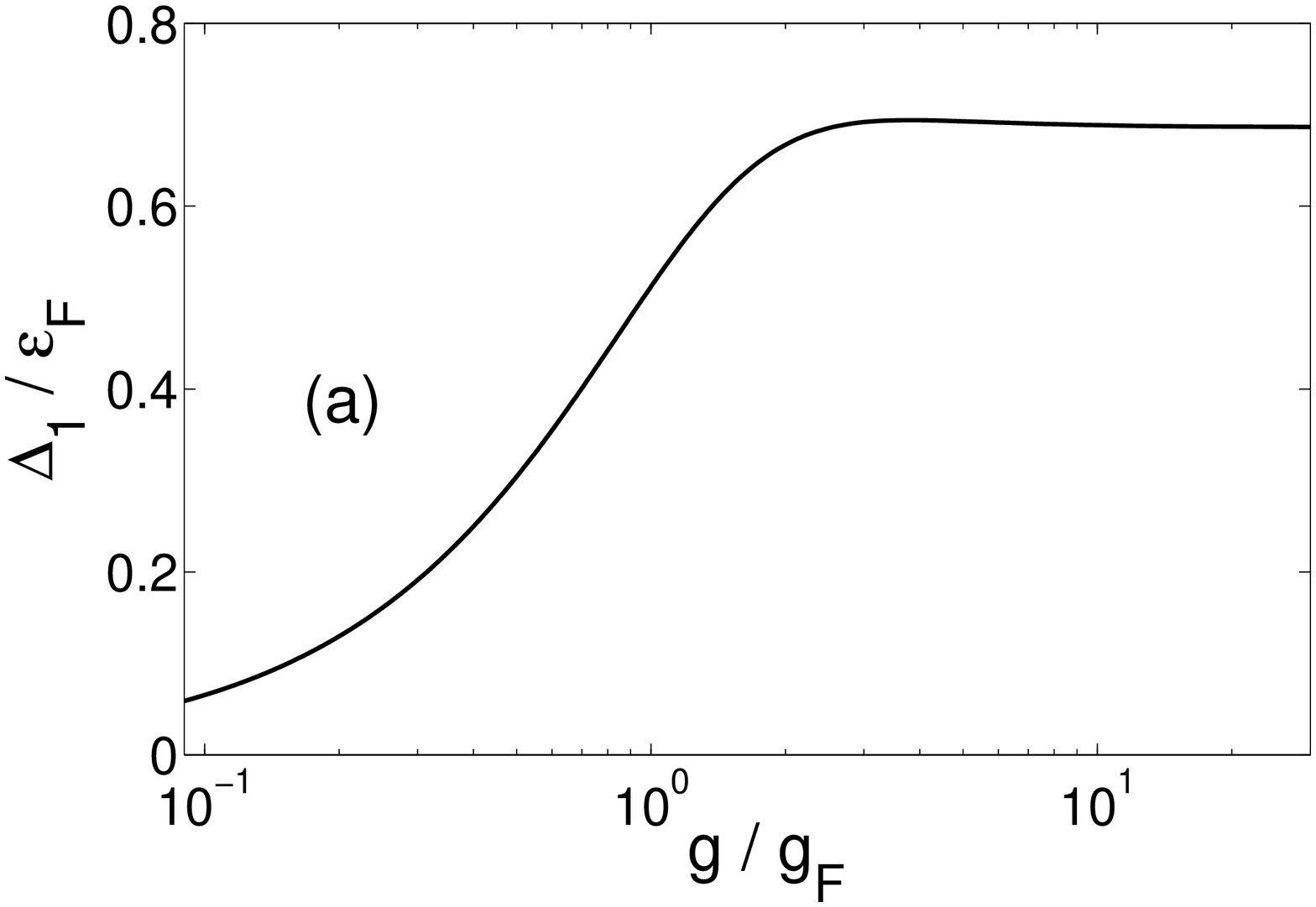}
\includegraphics[width=8cm]{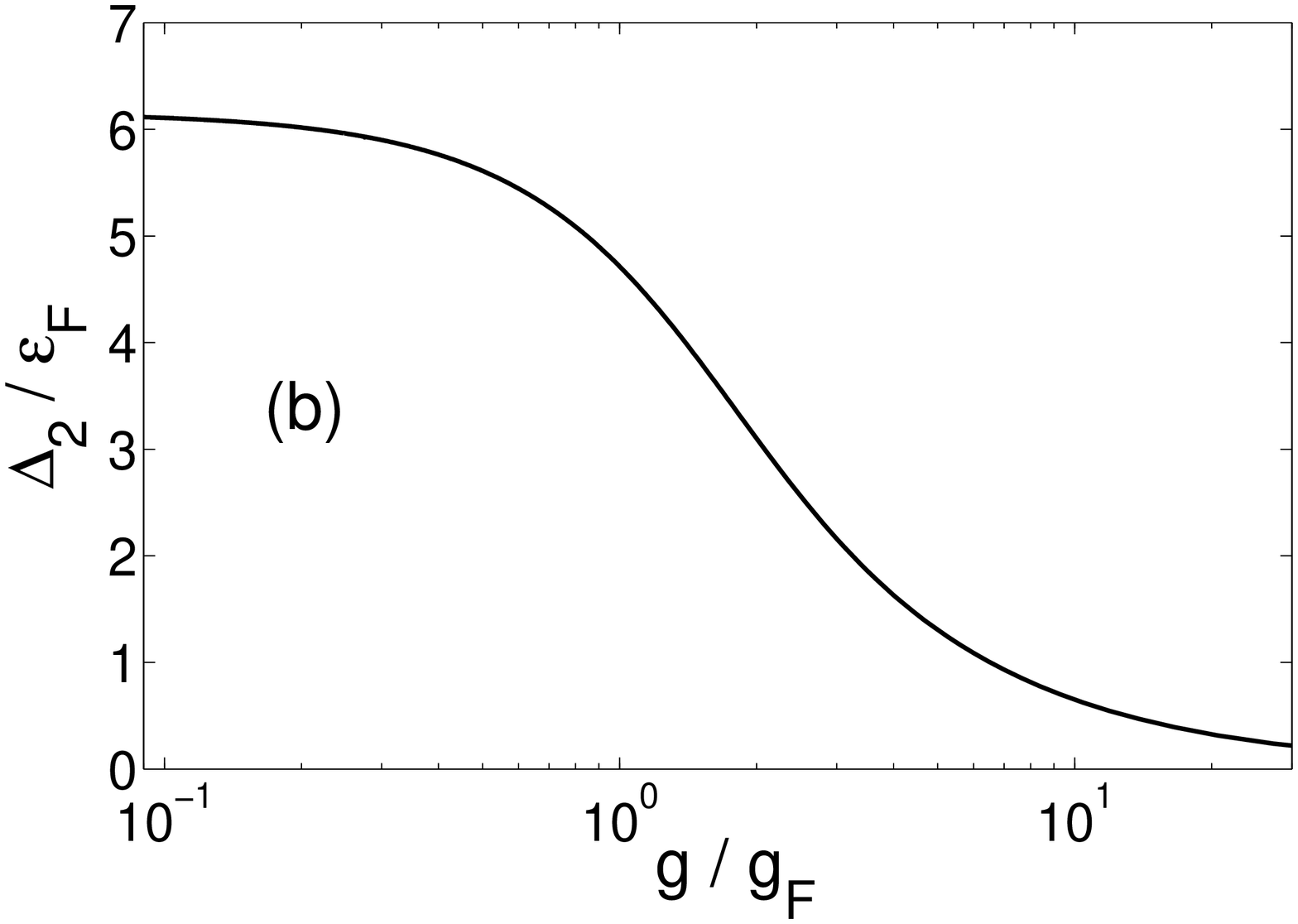}
\includegraphics[width=8cm]{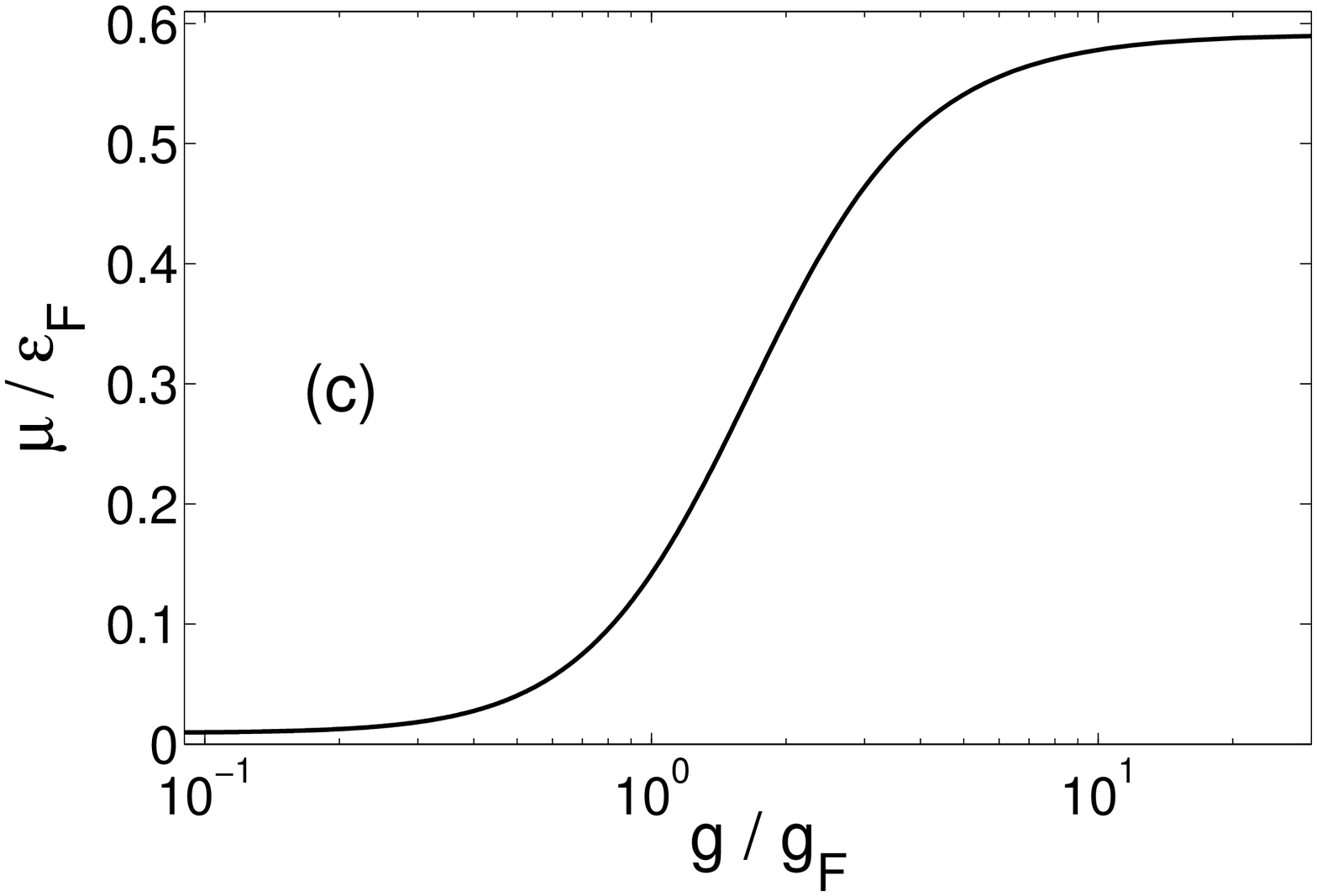}
\includegraphics[width=8cm]{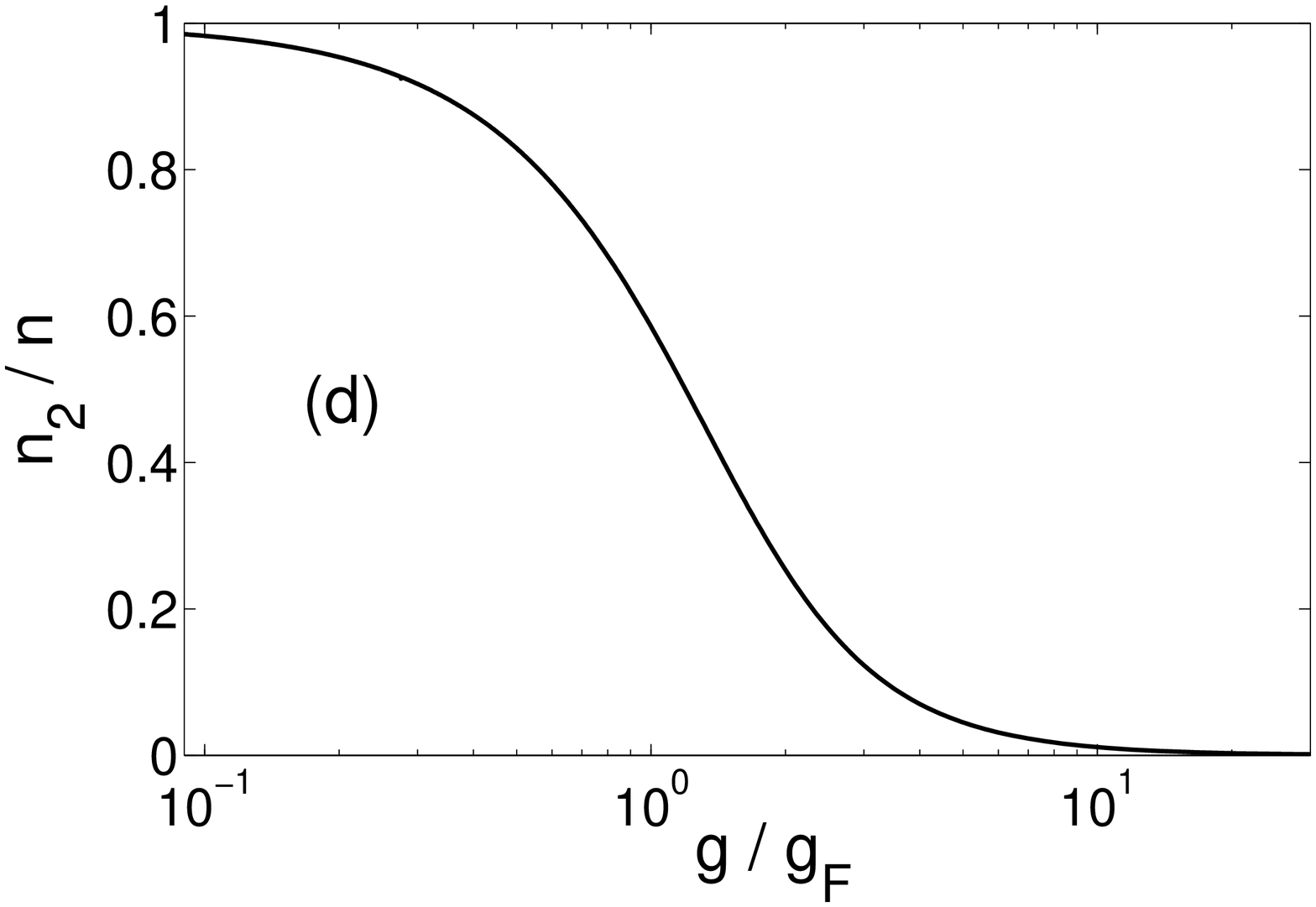}
\caption{(Color online) The dependence of the pairing gaps $\Delta_1$ (a) and $\Delta_2$ (b), the chemical potential $\mu$ (c), and the 
closed-channel fraction $n_2/n$ (d) on the coupling $g$ (scaled by $g_{\rm F}=\sqrt{2\pi k_{\rm F}}/M$) at the resonance ($\delta=0$). The coupling $g$ is determined by $g=U_{12}(M\varepsilon_0)^{3/4}/\sqrt{2\pi}$, with $\varepsilon_0=10^3\varepsilon_{\rm F}$. The background scattering length is set to be $k_{\rm F}a_{\rm bg}=0.1$. \label{fig2}}
\end{center}
\end{figure*}

To express $T_c$ in terms of the density $n$ or $\varepsilon_{\rm F}$, we need to solve the chemical potential $\mu$ through the number equation $n=n_1+n_2+n_{\rm g}$. The
mean-field contributions can be simplified as
\begin{eqnarray}
n_{\rm n}=2\sum_{\bf k}f(\varepsilon_{\bf k}-\mu_{\rm n}).
\end{eqnarray}
We have $n_2\simeq0$ for $T_c\ll\varepsilon_0$.  The fluctuation contribution $n_{\rm g}$ is given by $n_{\rm g}=-\partial\Omega_{\rm g}/\partial\mu$. For vanishing order parameters, the effective action ${\cal S}_{\rm g}$ can be simplified as
\begin{equation}
{\cal S}_{\rm g}=\sum_Q\phi^\dagger(-Q)\Gamma^{-1}(Q)\phi(Q).
\end{equation}
Here $\phi^\dagger(-Q)=(\phi_1^*(Q),\phi_2^*(Q))$ and the inverse boson propagator $\Gamma^{-1}(Q)$ becomes a $2\times2$ matrix,
\begin{equation}
\Gamma^{-1}(Q)=-\left(\begin{array}{cc} U_{11}& U_{12}\\ U_{21}& U_{22}\end{array}\right)^{-1}+
\left(\begin{array}{cc} \chi_1(Q)& 0\\ 0& \chi_2(Q)\end{array}\right),
\end{equation}
where the pairing susceptibilities $\chi_{\rm n}(Q)$ reads
\begin{eqnarray}
\chi_{\rm n}(Q)=\sum_{\bf k}\left[\frac{1-f(\xi_{{\rm n}{\bf k}+{\bf q}/2})-f(\xi_{{\rm n}{\bf k}-{\bf q}/2})}
{i\omega_\nu+2\mu_{\rm n}-\frac{{\bf q}^2}{4M}-2\varepsilon_{\bf k}}+\frac{1}{2\varepsilon_{\bf k}}\right].
\end{eqnarray}
We note that the superfluid transition temperature is also given by $\det\Gamma^{-1}(0,0)=0$, which is the generalized Thouless criterion for two-band systems. Finally, the contribution $\Omega_{\rm g}$ can be expressed as
\begin{eqnarray}
\Omega_{\rm g}=-\sum_{\bf q}\int_{-\infty}^\infty\frac{d\omega}{\pi}\frac{1}{e^{\beta\omega}-1}
\left[\delta_1(\omega,{\bf q})+\delta_2(\omega,{\bf q})\right],
\end{eqnarray}
where $\delta_{\rm n}(\omega,{\bf q})=-{\rm Im}\ln [\Gamma_{\rm n}^{-1}(\omega+i\epsilon,{\bf q})]$, with the two vertex functions given by
\begin{eqnarray}
&&\Gamma_1^{-1}(i\omega_\nu,{\bf q})=-\left[U_{11}+\frac{U_{12}^2\chi_2(Q)}{1-U_{22}\chi_2(Q)}\right]^{-1}+\chi_1(Q),\nonumber\\
&&\Gamma_2^{-1}(i\omega_\nu,{\bf q})=-\left(U_{22}-\frac{U_{12}^2}{U_{11}}\right)^{-1}+\chi_2(Q).
\end{eqnarray}
The first contribution corresponds to the usual Nozières-Schmitt-Rink approach with an energy-dependent scattering length \cite{NSR-narrow}. The second contribution can be attributed to the presence of the closed-channel. Actually, for $\varepsilon_0\rightarrow\infty$, we have
\begin{eqnarray}
&&\frac{U_{12}^2\chi_2(Q)}{1-U_{22}\chi_2(Q)}\simeq\frac{g^2}{i\omega_\nu-\frac{{\bf q}^2}{4M}+2\mu-\delta},\nonumber\\
&&\Gamma_2^{-1}(i\omega_\nu,{\bf q})\simeq-\alpha\left(i\omega_\nu-\frac{{\bf q}^2}{4M}+2\mu-\delta+\gamma B_\Delta\right).
\end{eqnarray}
We expect that the term $\gamma B_\Delta$ in $\Gamma_2^{-1}(i\omega_\nu,{\bf q})$ controls the closed-channel contribution. For broad resonance
with $\gamma B_\Delta\gg\varepsilon_{\rm F}$, this contribution can be safely neglected and we recover the single-channel description.

\section {Dilute Limit: $\varepsilon_0\rightarrow\infty$}\label{s5}
In this section, we study the dilute limit of the two-band theory. The dilute limit means $\varepsilon_0/\varepsilon_{\rm F}\rightarrow\infty$.
The closed-channel binding energy $\varepsilon_0$ is equal to the Zeeman energy splitting at the resonance $B=B_0$ \cite{Bruun}. Considering the resonance occurs at high magnetic field, we estimate $\varepsilon_0\sim\gamma B_0$. For the $^6$Li atom, its broad resonance and the narrow resonance occur at $B=834.1$ G and $B=543.25$ G, respectively. The typical density of atoms realized in current experiments is $10^{13}-10^{14}$ cm$^{-3}$. Therefore, we estimate that the ratio $\varepsilon_0/\varepsilon_{\rm F}$ is of order $10^3$ in current experimental systems, which satisfies well the dilute condition $\varepsilon_{\rm F}\ll\varepsilon_0$.

\begin{figure*}[!htb]
\begin{center}
\includegraphics[width=8cm]{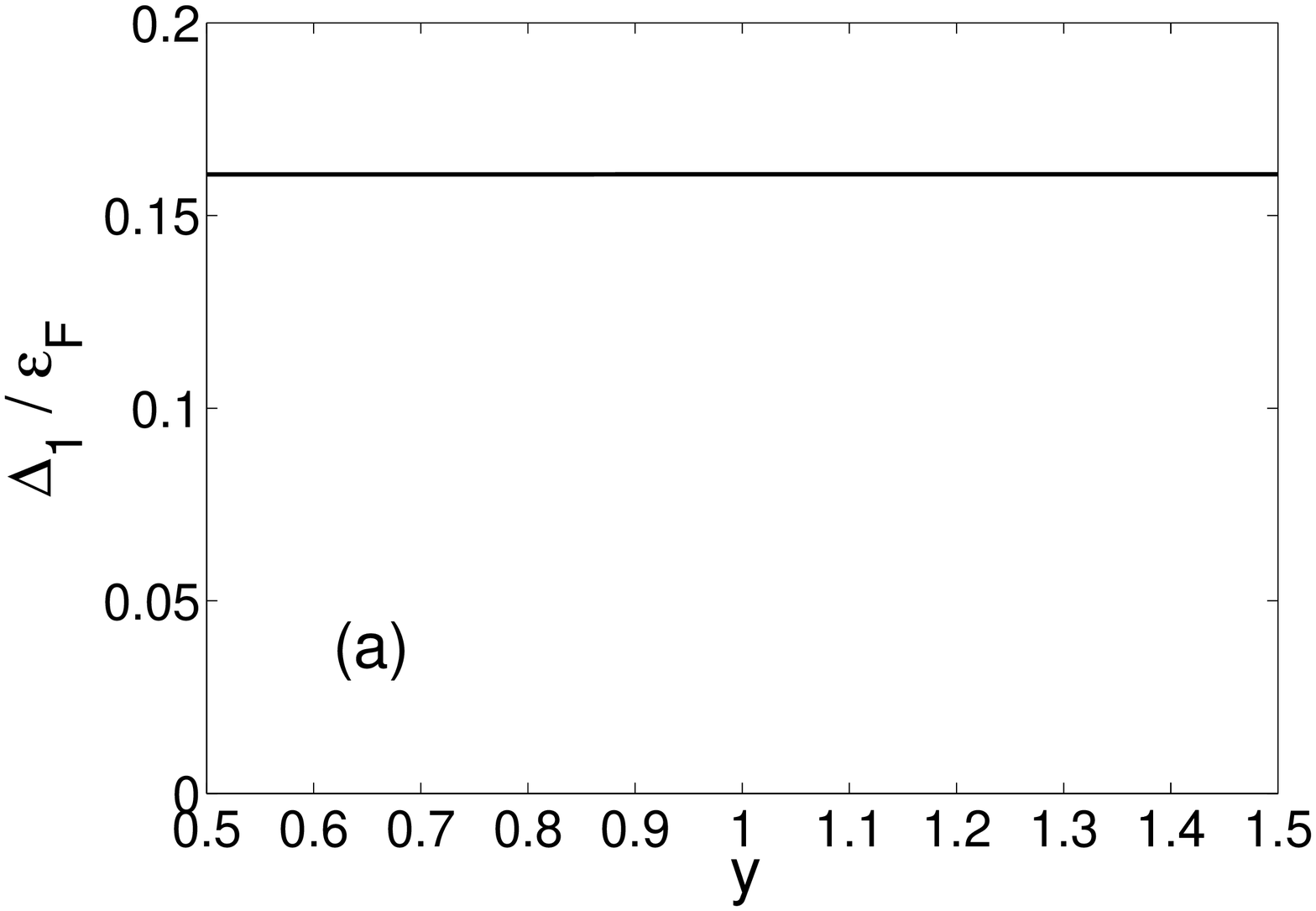}
\includegraphics[width=8cm]{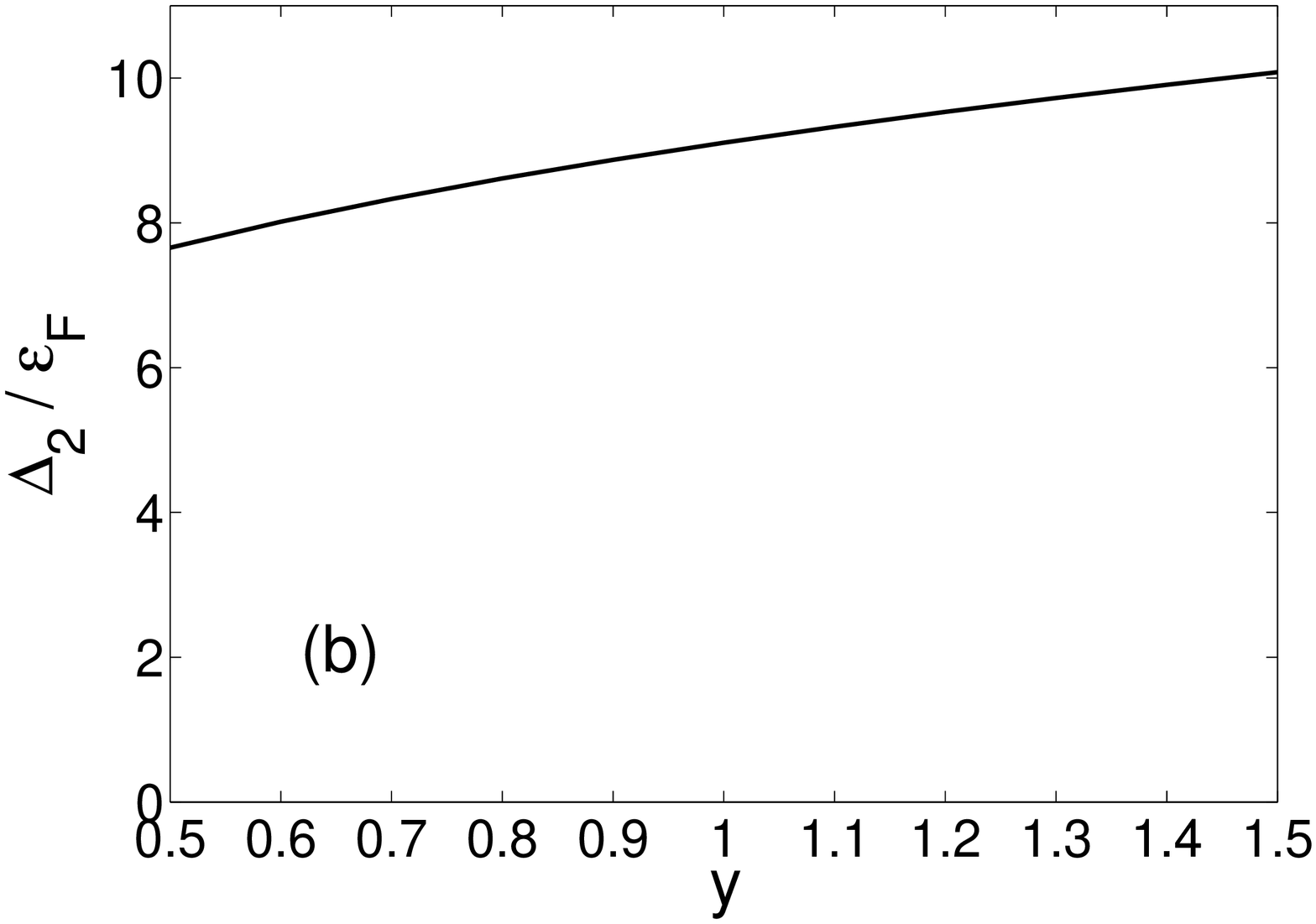}
\includegraphics[width=8cm]{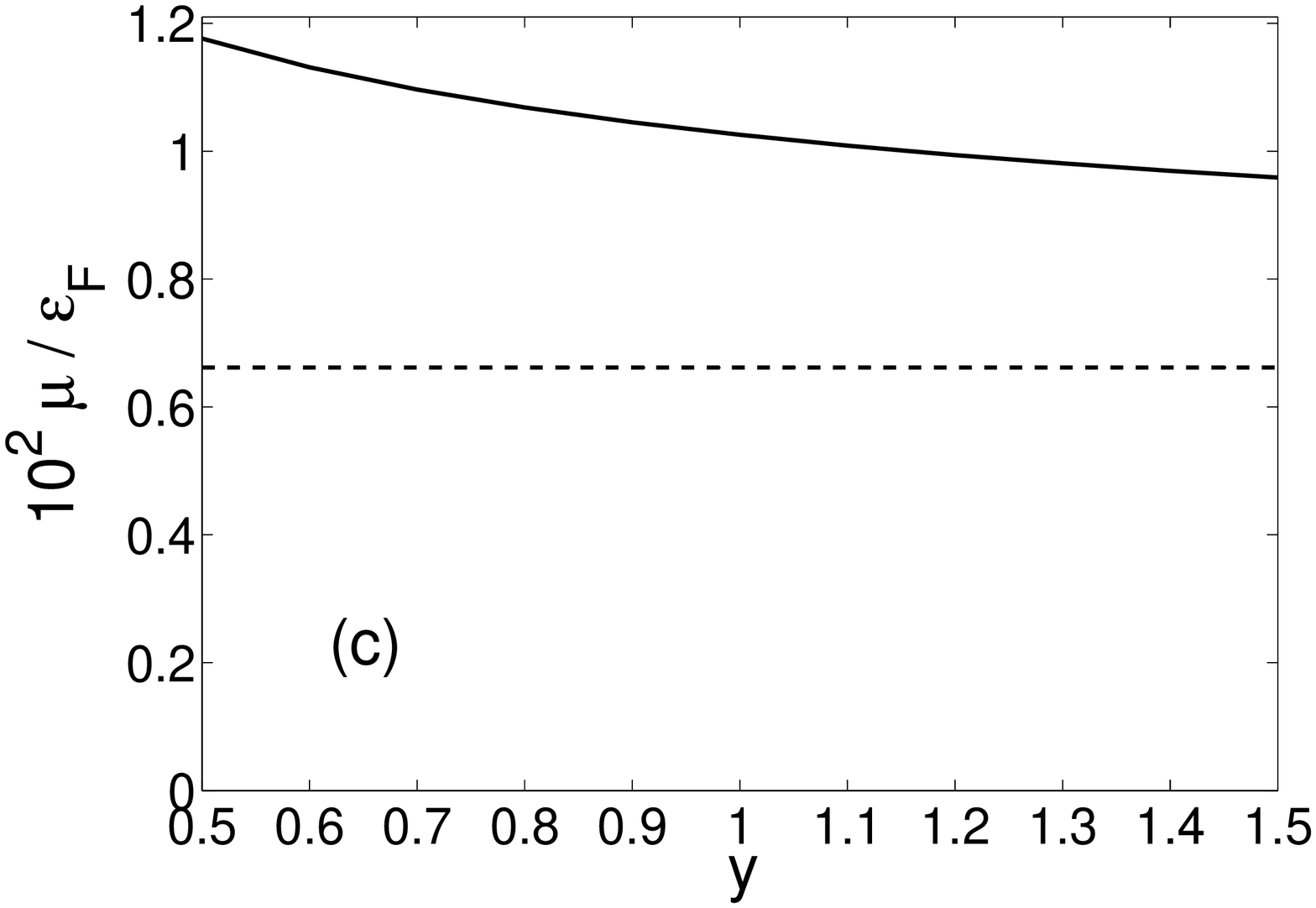}
\includegraphics[width=8cm]{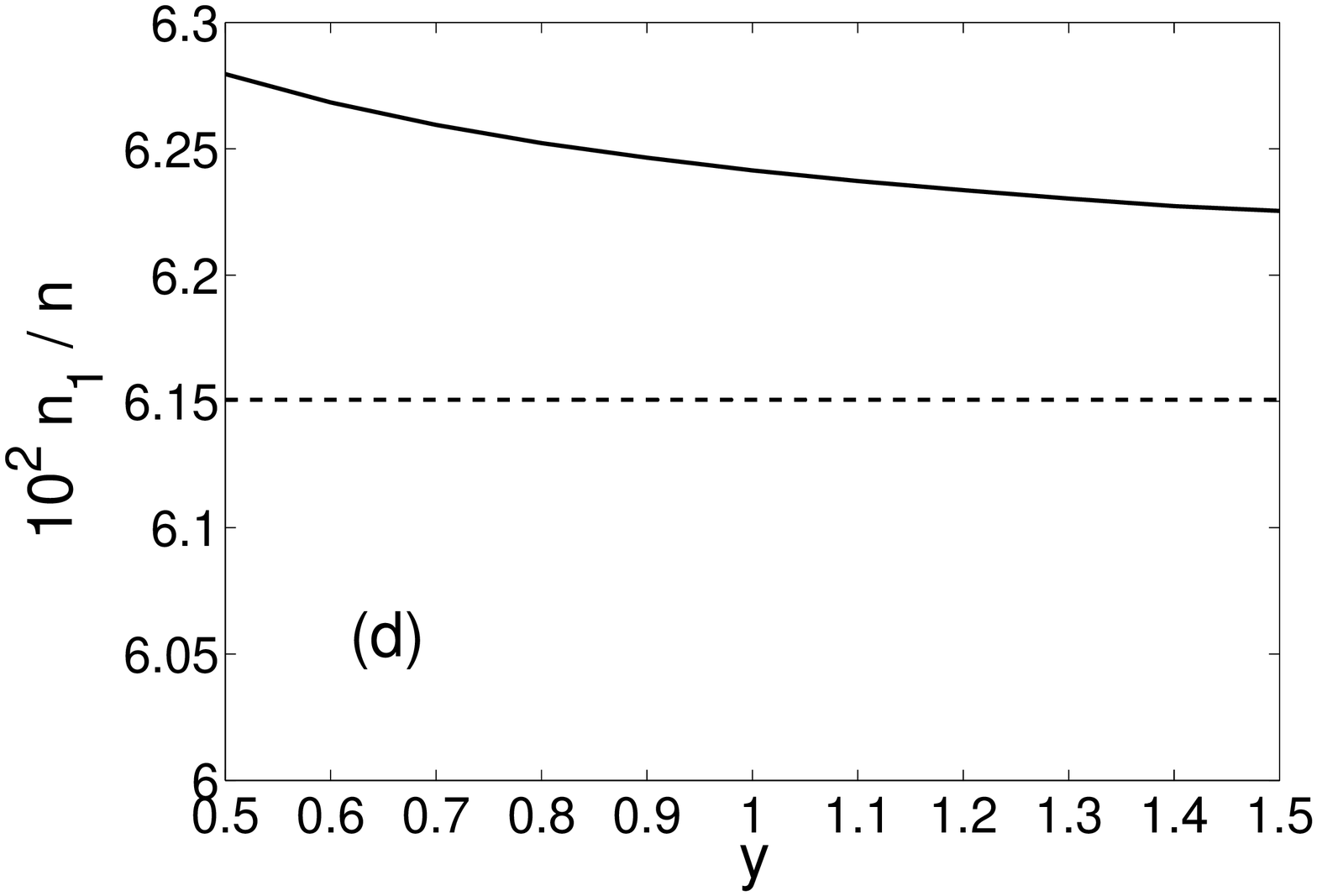}
\caption{The dependence of the pairing gaps $\Delta_1$ (a) and $\Delta_2$ (b), the chemical potential $\mu$ (c), and the open-channel fraction $n_1/n$ (d) on the parameter $y=\varepsilon_0/(\gamma B_0)$ at the resonance. In the calculations we set the density parameter $x=\varepsilon_{\rm F}/(\gamma B_\Delta)=1$, which corresponds to $g/g_{\rm F}=0.25$ or $k_{\rm F}r_{\rm eff}=-63$. The dashed lines are the predictions from atom-molecule theory, which coincides with the dilute limit ($\varepsilon_0/\varepsilon_{\rm F}\rightarrow\infty$) of the two-band theory. \label{fig3}}
\end{center}
\end{figure*}

For the sake of simplicity, we focus on the zero-temperature case and employ the mean-field theory. We show that in the mean-field approximation, 
the predictions from the two-band theory become essentially the same as the atom-molecule model in the dilute limit.
The coupled gap equations for the pairing gaps $\Delta_1$ and $\Delta_2$ can be expressed as
\begin{eqnarray}
&&\frac{1}{U_{11}+{\cal C}(\Delta_2)}=F_1(\Delta_1),\nonumber\\
&&\frac{\Delta_2}{\Delta_1}=\frac{{\cal C}(\Delta_2)}{U_{11}+{\cal C}(\Delta_2)}\frac{1}{U_{12}F_2(\Delta_2)},
\end{eqnarray}
where ${\cal C}(\Delta_2)$ is defined as
\begin{equation}
{\cal C}(\Delta_2)=\frac{U_{12}^2F_2(\Delta_2)}{1-U_{22}F_2(\Delta_2)}.
\end{equation}
The quantity ${\cal C}$ shows explicitly the resonance effect on the open channel. Note that $\Delta_1$ and $\Delta_2$ are both complex quantities. Without loss of generality, we set $\Delta_1$ to be real and positive. From the second equation we find that $\Delta_2$ is also real.

In the dilute limit $\varepsilon_{\rm F}\ll\varepsilon_0$, we expect that $|\delta|,|\mu|\ll\varepsilon_0$ near the FR. Meanwhile, we also assume that $|\Delta_2|\ll\varepsilon_0$. While this is not evident at present, we prove it self-consistently. Therefore, for $\varepsilon_0\rightarrow\infty$, the function $F_2(\Delta_2)$ asymptotically behaves as
\begin{eqnarray}
F_2(\Delta_2)=\frac{M}{4\pi}\sqrt{M(\varepsilon_0+\delta-2\mu)}\left[1+O\left(\alpha^2\right)\right],
\end{eqnarray}
where $\alpha=|\Delta_2|/\varepsilon_0$. Then we obtain
\begin{eqnarray}
\lim_{\varepsilon_0\rightarrow\infty}{\cal C}(\Delta_2)
&=&\lim_{\varepsilon_0\rightarrow\infty}\frac{MU_{12}^2}{4\pi}\frac{\sqrt{M(\varepsilon_0+\delta-2\mu)}}
{1-\frac{\sqrt{M(\varepsilon_0+\delta-2\mu)}}{\sqrt{M\varepsilon_0}}}\nonumber\\
&=&\frac{g^2}{2\mu-\delta}\equiv{\cal C}_\infty,
\end{eqnarray}
where the atom-molecule coupling $g$ is given by
\begin{eqnarray}
g=\lim_{U_{12}\rightarrow0}\lim_{\varepsilon_0\rightarrow\infty}U_{12}\sqrt{\frac{(M\varepsilon_0)^{3/2}}{2\pi}}.
\end{eqnarray}
Thus the first gap equation becomes essentially the same as the gap equation of the atom-molecule model.

\begin{figure*}[!htb]
\begin{center}
\includegraphics[width=8cm]{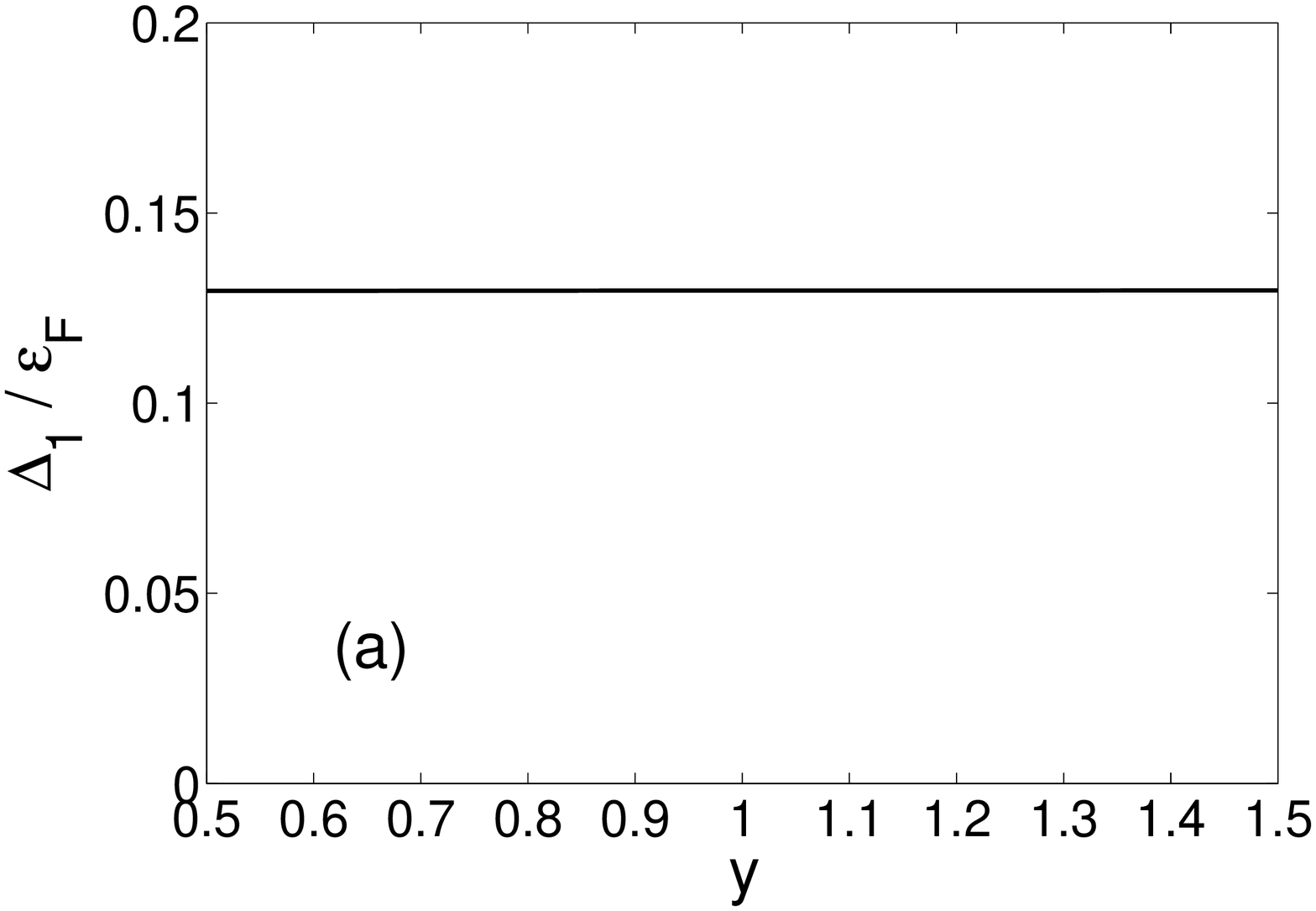}
\includegraphics[width=8cm]{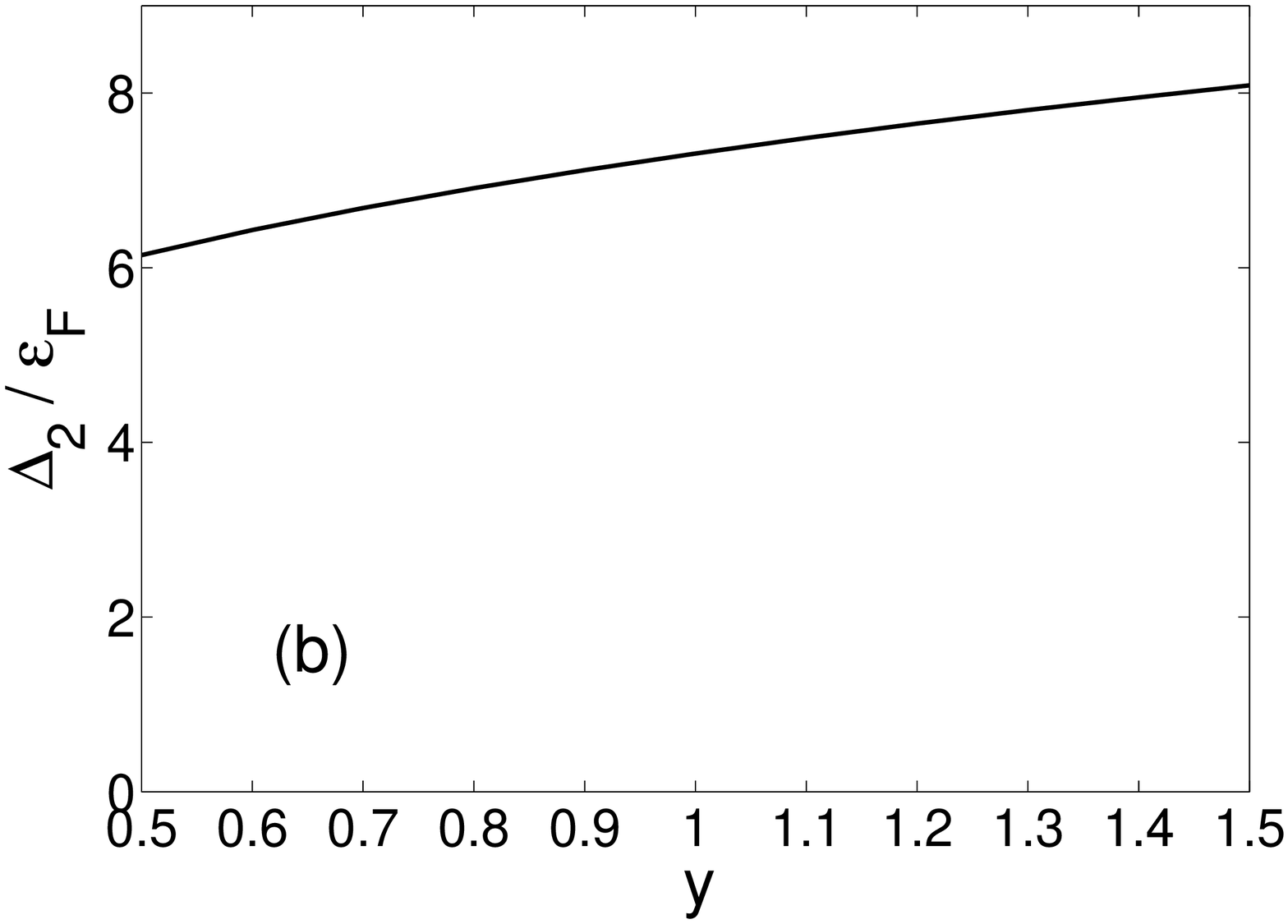}
\includegraphics[width=8cm]{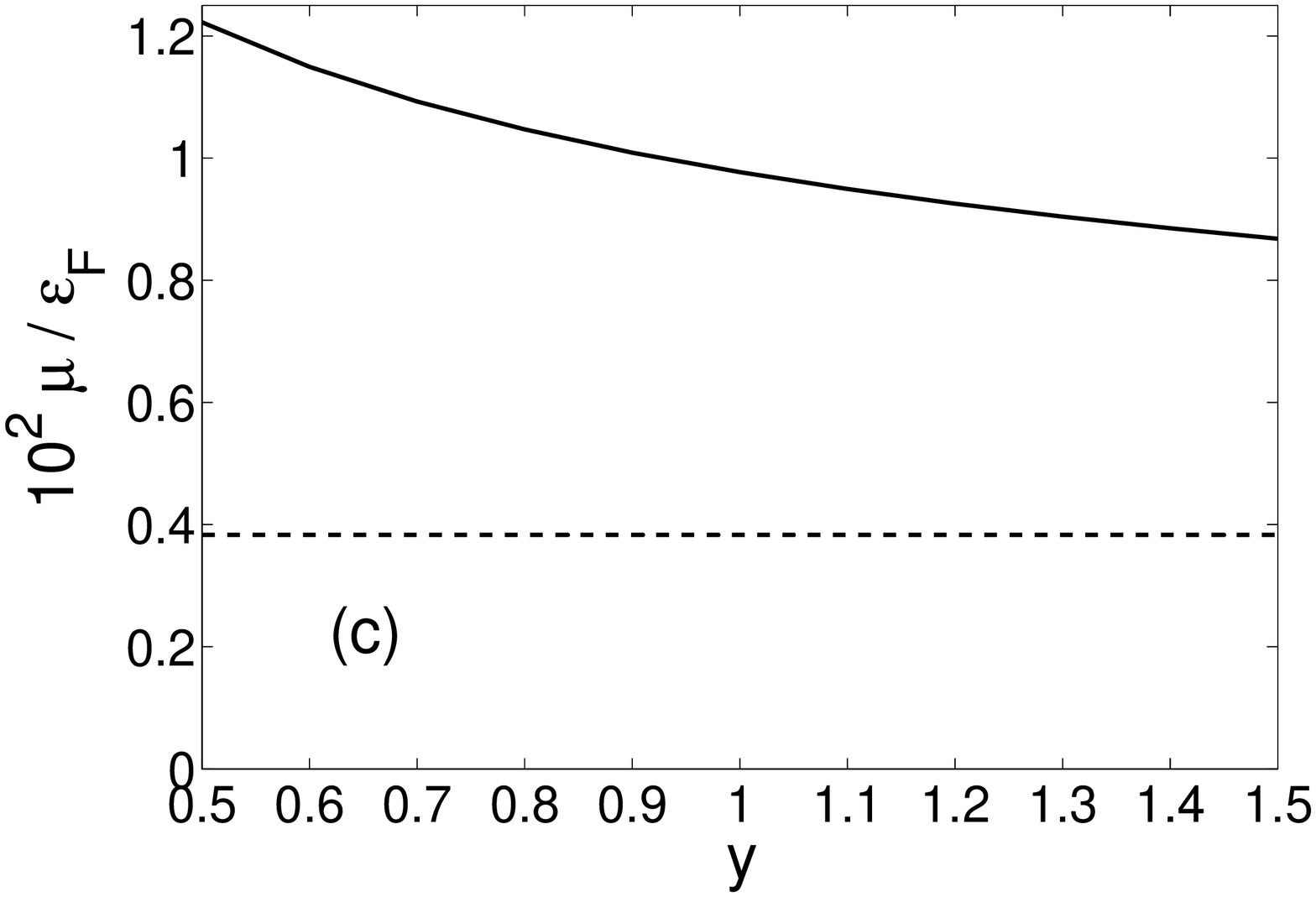}
\includegraphics[width=8cm]{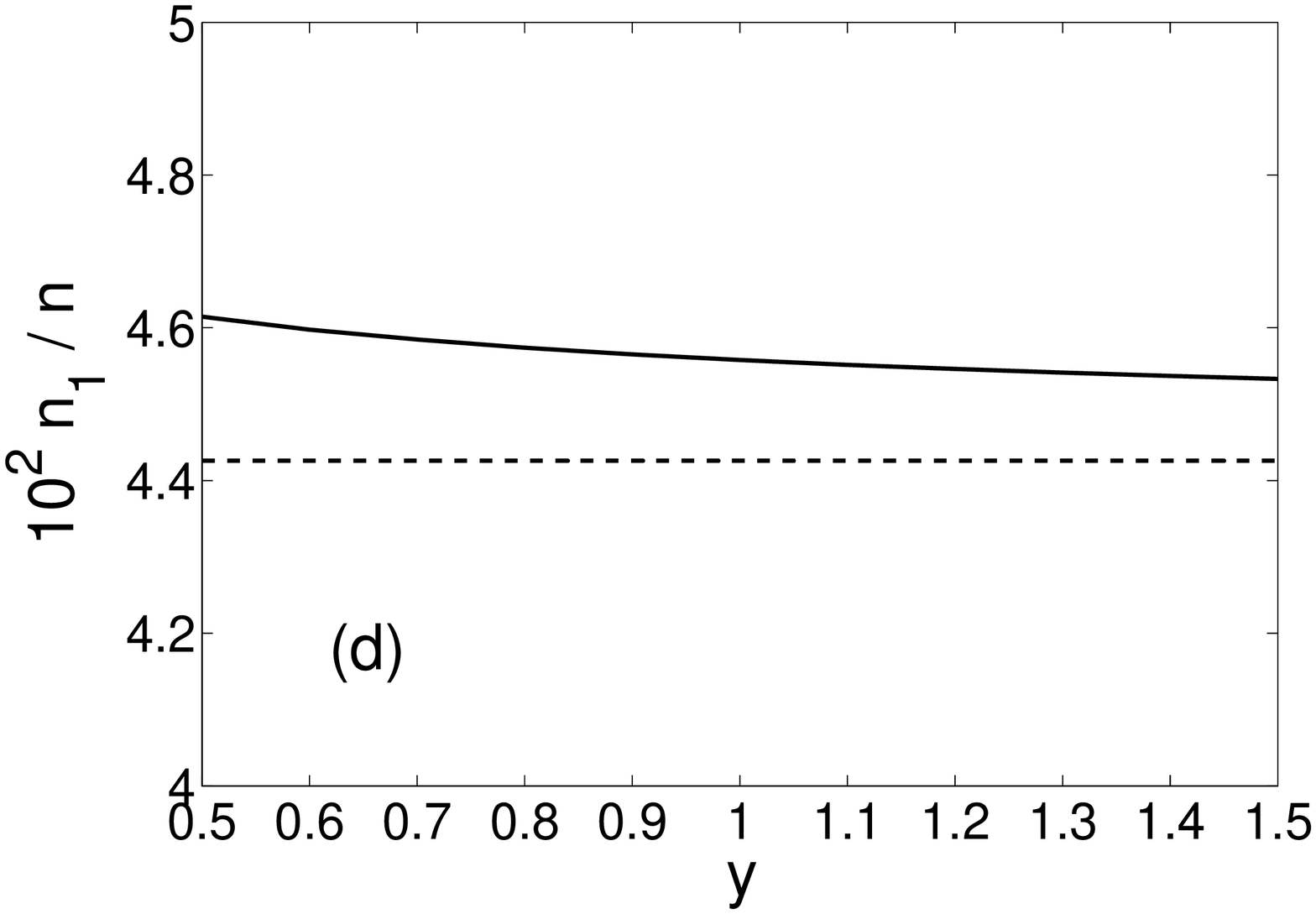}
\caption{The dependence of the pairing gaps $\Delta_1$ (a) and $\Delta_2$ (b), the chemical potential $\mu$ (c), and the open-channel fraction $n_1/n$ (d) on the parameter $y=\varepsilon_0/(\gamma B_0)$ at the resonance. In the calculations we set the density parameter $x=\varepsilon_{\rm F}/(\gamma B_\Delta)=2.5$, which corresponds to $g/g_{\rm F}=0.2$ or $k_{\rm F}r_{\rm eff}=-100$. The dashed lines are the predictions from atom-molecule theory, which coincides with the dilute limit ($\varepsilon_0/\varepsilon_{\rm F}\rightarrow\infty$) of the two-band theory. \label{fig4}}
\end{center}
\end{figure*}

For the second equation, in the dilute limit we have
\begin{eqnarray}\label{gapratio}
\frac{\Delta_2}{\Delta_1}=\frac{{\cal C}_\infty}{U_{11}+{\cal C}_\infty}
\frac{2\sqrt{2\pi}}{gM}(M\varepsilon_0)^{1/4}\left[1+O\left(\alpha^2\right)\right],
\end{eqnarray}
For $\varepsilon_0\rightarrow\infty$, we have $\Delta_2\rightarrow\infty$ but $\Delta_2/\varepsilon_0\rightarrow0$. Using this result, we can
simplify the number equation. In the mean-field theory, the number equation is given by $n=n_1+n_2$, where $n_1$ and $n_2$ are the contributions from
the open channel and closed channel, respectively. Since $\Delta_2/\varepsilon_0\rightarrow0$ for $\varepsilon_0\rightarrow\infty$, we can expand the closed-channel contribution $n_2$ in powers of $|\Delta_2|/|\mu_2|$. Therefore, the number density of the closed channel asymptotically behaves as
\begin{eqnarray}
n_2=\frac{\pi}{8}\frac{|\Delta_2|^2}{|\mu_2|^2}\frac{(2M|\mu_2|)^{3/2}}{2\pi^2}\left[1+O\left(\alpha^2\right)\right].
\end{eqnarray}
Using the relation (\ref{gapratio}), we obtain
\begin{eqnarray}
\lim_{\varepsilon_0\rightarrow\infty}n_2=\frac{2|\Delta_1|^2}{g^2}\left(\frac{{\cal C}_\infty}{U_{11}+{\cal C}_\infty}\right)^2.
\end{eqnarray}
This is the same as the closed-channel contribution $n_{\rm m}$ in the atom-molecule model. Therefore, in the dilute limit, the pairing gap $\Delta_2$ can be eliminated and the predictions become the same as the atom-molecule model.

For the grand potential $\Omega_0$, by using the asymptotical behavior
\begin{eqnarray}
&&\sum_{\bf k}\left(\xi_{2\bf k}-E_{2\bf k}+\frac{|\Delta_2|^2}{2\varepsilon_{\bf k}}\right)\nonumber\\
&=&\frac{M|\Delta_2|^2}{4\pi}\sqrt{M(\varepsilon_0+\delta-2\mu)}\left[1+O\left(\alpha^2\right)\right]
\end{eqnarray}
and the relation (102), we can show that it recovers the result (30) of the atom-molecule model.

In realistic experimental systems, the ratio $\varepsilon_0/\varepsilon_{\rm F}$ is large but finite. We expect that the predictions from the two-band theory agree with the atom-molecule model in addition to a tiny correction, which should be of order $O(\varepsilon_F/\varepsilon_0)$. In Fig. \ref{fig2} we show the evolution of the pairing gaps, the chemical potential, and the closed-channel fraction with the interchannel coupling $U_{12}$ at the resonance for $\varepsilon_0=10^3\varepsilon_{\rm F}$ and $k_{\rm F}a_{\rm bg}=0.1$. In the plots, we have also used the atom-molecule coupling $g$, which is determined by $g=U_{12}(M\varepsilon_0)^{3/4}/\sqrt{2\pi}$ and scaled by $g_{\rm F}=\sqrt{2\pi k_{\rm F}}/M$.
The effective range parameter $k_{\rm F}r_{\rm eff}$ is related to the coupling as
\begin{eqnarray}
k_{\rm F}r_{\rm eff}=-4\left(\frac{g}{g_{\rm F}}\right)^{-2}.
\end{eqnarray}
For sufficiently large coupling $U_{12}$ or $g$, where $|k_{\rm F}r_{\rm eff}|\ll1$, the system enters the universal regime. In this regime we have $n_2\rightarrow0$ and hence the open channel dominates. The open-channel pairing gap $\Delta_1$ and the chemical potential $\mu$ agree with the universal values $\Delta_1=0.6864\varepsilon_{\rm F}$ and $\mu=0.5906\varepsilon_{\rm F}$ from the single-channel model. From the numerical results shown in Fig. \ref{fig2}, we find that the crossover from the broad to narrow resonances occurs at $|k_{\rm F}r_{\rm eff}|\sim1$.
We have also compared the results with the predictions from the atom-molecule model with the same coupling $g$.
For $\varepsilon_0=10^3\varepsilon_{\rm F}$, we find that the two-band predictions already agree well with the predictions from the atom-molecule model. For broad and moderate resonances, the corrections to the pairing gap $\Delta_1$ and the chemical potential $\mu$ are tiny. Our results agree with a recent multichannel approach to the pairing in atomic Fermi gases where the open and closed channels have one hyperfine state in common \cite{Zhu}. On the other hand, we find from our numerical analysis that the correction to the (dimensionless) chemical potential $\mu/\varepsilon_{\rm F}$ is generally of order $O(\varepsilon_{\rm F}/\varepsilon_0)$ for $\varepsilon_{\rm F}/\varepsilon_0\sim10^{-3}$.
We notice that $\mu/\varepsilon_{\rm F}\rightarrow0$ in the narrow resonance limit. Therefore, for extremely narrow resonance, this tiny correction may become significant because the chemical potential itself is also tiny. We focus on the extremely narrow resonance in the next section. On the other hand, at high density where $\varepsilon_{\rm F}\sim\varepsilon_0$, medium effects on the closed channel become significant and the atom-molecule model becomes invalid. Unfortunately, this high-density regime is not accessible in current experiments of atomic Fermi gases.

\section {Extremely Narrow resonance}\label{s6}

As we mentioned above, the tiny correction due to nonvanishing $\varepsilon_{\rm F}/\varepsilon_0$ may become remarkable for extremely narrow resonance with $g/g_{\rm F}\ll1$ or $|k_{\rm F}r_{\rm eff}|\gg1$, because the chemical potential $\mu/\varepsilon_{\rm F}$ itself becomes comparable
with this tiny correction. An intuitive picture is that, for extremely narrow resonance, the closed-channel dominates and the system can be regarded as a BEC of the closed-channel molecules. In the atom-molecule model, the closed-channel molecules are treated as noninteracting point bosons. However, in the present two-band theory, the closed-channel molecules are treated as composite bosons and theirs interactions are automatically taken into account. The leading correction is the two-body boson-boson interaction with a scattering length $a_{\rm m}$, as we have shown in Eq. (69). The interaction parameter $k_{\rm F}a_{\rm m}$ reads
\begin{eqnarray}
k_{\rm F}a_{\rm m}=\frac{2k_{\rm F}}{\sqrt{M\varepsilon_0}}=2\sqrt{\frac{2\varepsilon_{\rm F}}{\varepsilon_0}}.
\end{eqnarray}
For realistic value $\varepsilon_{\rm F}/\varepsilon_0\sim10^{-3}$, the above interaction parameter is generally of order $0.1$. Therefore, for extremely narrow resonance, the boson-boson interaction can lead to remarkable correction to the chemical potential and the equations of state.

\begin{figure}[!htb]
\begin{center}
\includegraphics[width=9cm]{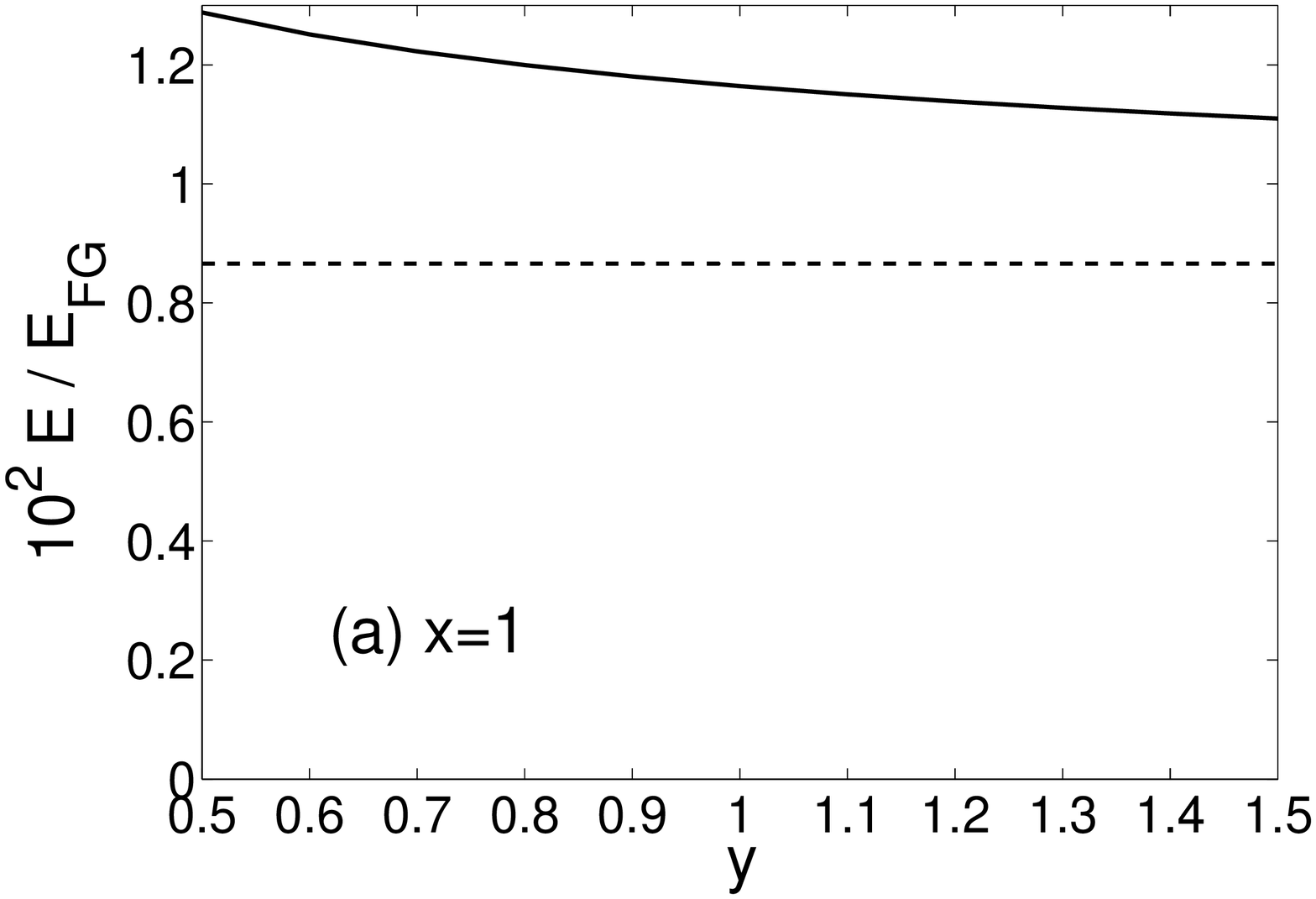}
\includegraphics[width=9cm]{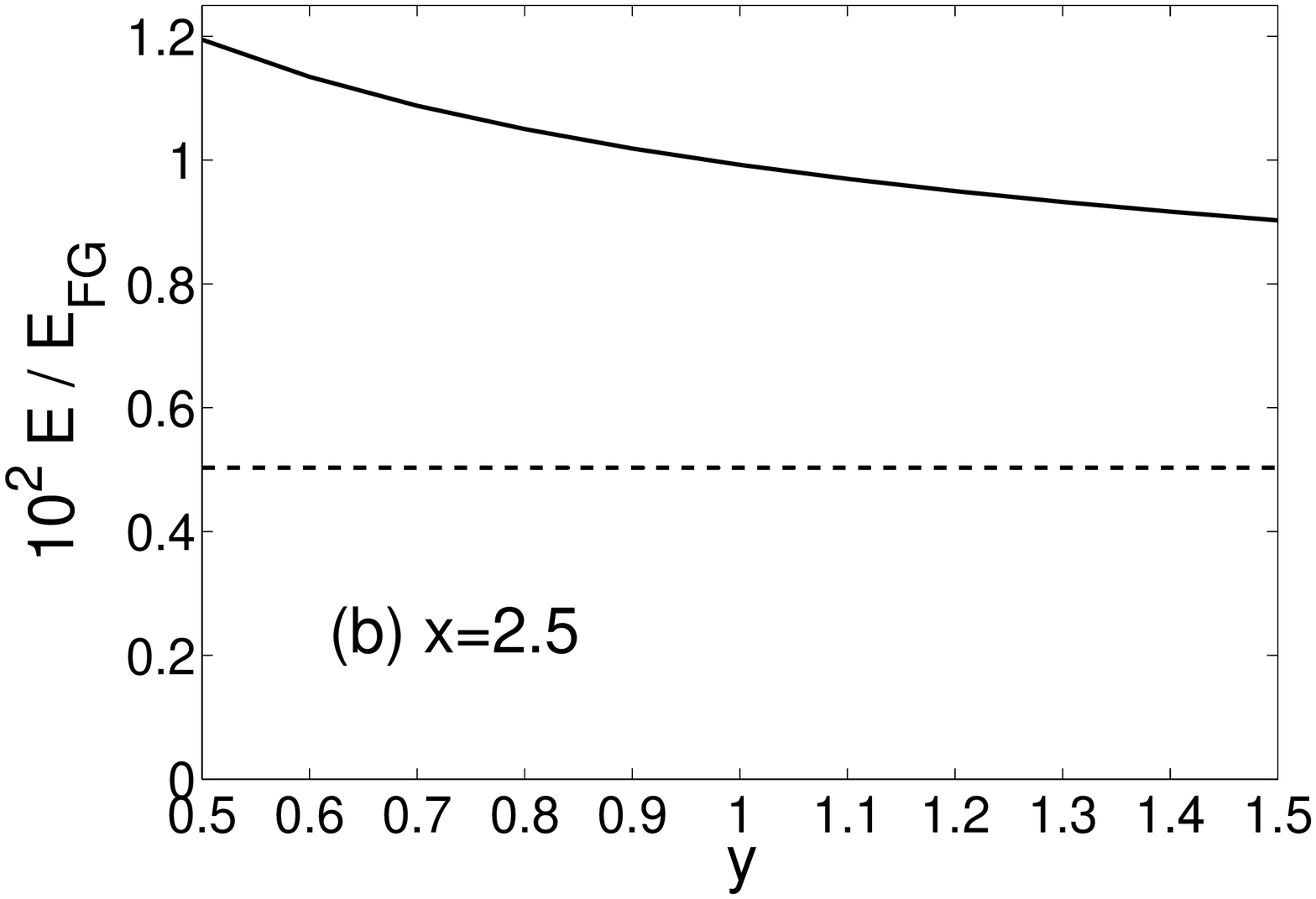}
\caption{The dependence of the energy of the resonant superfluid on the parameter $y=\varepsilon_0/(\gamma B_0)$ for $x=1$ (a) and $x=2.5$ (b). The energy is scaled by the energy of the noninteracting two-component Fermi gas, $E_{\rm FG}=\frac{3}{5}N\varepsilon_{\rm F}$. The dashed lines are predictions from atom-molecule theory, corresponding to the dilute limit $\varepsilon_0/\varepsilon_{\rm F}\rightarrow\infty$. \label{fig5}}
\end{center}
\end{figure}

In the final part of this work, we apply the two-band theory to study the narrow resonance of $^6$Li atoms at $B_0=543.25$ G. The resonance width and the background scattering length have been measured to be $B_\Delta=0.1$ G and $a_{\rm bg}=61.6a_{\rm B}$ \cite{Expnarrow}. For convenience, we define two parameters,
\begin{eqnarray}
a=\frac{B_\Delta}{B_0},\ \ \ \ \ \ b=\frac{\gamma B_\Delta}{\varepsilon_{\rm bg}},
\end{eqnarray}
where $\varepsilon_{\rm bg}=1/(Ma_{\rm bg}^2)$ is the energy associated with the background scattering length. We also define the following two variables:
\begin{eqnarray}
x=\frac{\varepsilon_{\rm F}}{\gamma B_\Delta},\ \ \ \ \ \ y=\frac{\varepsilon_0}{\gamma B_0}.
\end{eqnarray}
For the typical densities realized in recent experiments \cite{Expnarrow}, we have $\varepsilon_{\rm F}\sim\gamma B_\Delta$ and, hence, $x\sim1$. The binding energy $\varepsilon_0$ is a parameter in the two-band model and so far cannot be extracted precisely. However, it is reasonable to estimate $\varepsilon_0\sim\gamma B_0$ and, hence, $y\sim1$. From the above parameters we obtain
\begin{eqnarray}
&&k_{\rm F}a_{\rm bg}=\sqrt{2bx},\ \ \ \ \ \ \ \ \ \
\frac{\varepsilon_0}{\varepsilon_{\rm F}}=\frac{y}{ax},\nonumber\\
&&\frac{U_{12}}{U_{\rm F}}=\left(\frac{a^3bx^2}{y^3}\right)^{1/4},\ \ \
\frac{g}{g_{\rm F}}=\left(\frac{2b}{x}\right)^{1/4},
\end{eqnarray}
where $U_{\rm F}=4\pi/(Mk_{\rm F})$. From the measurements we have $b=2\times10^{-3}$, which means that this resonance is extremely narrow. In the following, we consider two typical densities, $x=1$ and $x=2.5$, which correspond to two effective range parameters $k_{\rm F}r_{\rm eff}=-63$ and
$k_{\rm F}r_{\rm eff}=-100$, respectively.

In Figs. \ref{fig3} and \ref{fig4}, we show the dependence of the pairing gaps, the chemical potential, and the open-channel fraction on the parameter $y$ in the range $0.5<y<1.5$ at the resonance ($\delta=0$) for two typical densities $x=1$ and $x=2.5$. At both densities, the finite-$\varepsilon_0$ corrections to the open-channel pairing gap $\Delta_1$ and the open-channel fraction $n_1/n$ are relatively small. However, the correction to the chemical potential $\mu$ becomes significant since the chemical potential itself is already very small for such an extremely narrow resonance. In Fig. \ref{fig5} we also show the energy of the resonant superfluid at two typical densities $x=1$ and $x=2.5$. For reasonable values of the closed-channel binding energy, i.e., $0.5<y<1.5$, we find that the chemical potential and the energy predicted from the two-band theory deviate significantly from those of the atom-molecule model predictions. For smaller $y$, we find that the deviation is larger. This can be understood by the fact that the boson-boson interaction parameter $k_{\rm F}a_{\rm m}$ becomes larger for smaller $y$. The atom-molecule model predictions correspond to the limit $y\rightarrow\infty$ of the two-band theory, which indicates vanishing boson-boson interaction $k_{\rm F}a_{\rm m}\rightarrow0$. However, for extremely narrow resonance, the convergence to the atom-molecule theory is very slow. From a numerical analysis, we find that the two-band theory predictions converge to the results from the atom-molecule theory at $y\sim10^4$.

\section {Summary and outlook}\label{s7}

In summary, we have shown that a simple two-band theory can describe the resonant superfluidity in atomic Fermi gases. The atom-molecule model
can be viewed as a low-energy effective theory of the two-band model in the limit $\varepsilon_0\rightarrow\infty$ and $U_{12}\rightarrow0$, while keeping the phenomenological atom-molecule coupling $g$ finite. Explicitly, the atom-molecule coupling $g$ is related to the microscopic parameters as $g=U_{12}(M\varepsilon_0)^{3/4}/\sqrt{2\pi}$. The two-band description of resonant superfluidity is in analogy to the BCS theory of
two-band superconductors. The closed-channel binding energy $\varepsilon_0$ provides a large band offset, which is automatically sent to infinity
in the atom-molecule model. In the dilute limit $\varepsilon_{\rm F}/\varepsilon_0\rightarrow0$, we find that the two-band theory reproduces
precisely the atom-molecule theory. Since the physical results do not depend on the details of the microscopic interaction potential
$V(|{\bf r}-{\bf r}^\prime|)$, the simple two-band model could be a feasible model for future Monte Carlo simulation of atomic Fermi gases across narrow FRs.

In realistic experimental systems, the ratio $\varepsilon_{\rm F}/\varepsilon_0$ is small but finite. The correction due to this small ratio physically corresponds to the effect of boson-boson interaction in the closed channel. For broad and moderate resonances, such correction is relatively small and thus not important. However, for extremely narrow resonance such as the resonance of $^6$Li at $B=543.25$ G, the correction becomes significant. The correction may also be important for the stability of the homogeneous polarized superfluid state against phase separation
for population imbalanced systems ($n_{1\uparrow}\neq n_{1\downarrow}$). A recent study of the polaron problem in highly polarized Fermi gases
across a narrow FR indicates that the highly polarized mixture can be stable against phase separation if the value
$k_{\rm F}a_{\rm m}$ is nonvanishing \cite{Zhai}, where $a_{\rm m}$ is the molecule-molecule scattering length in the closed channel. On the other hand, it has been shown that the polarized superfluid state can be stable against phase separation in two-band Fermi superfluids \cite{He}. Therefore, it is interesting to apply the two-band theory to study the possibility of a stable polarized superfluid state across a narrow FR.

\emph{Acknowledgments} --- The work of L. H. was supported by the U. S. Department of Energy Nuclear Physics Office (Contract No. DE-AC02-05CH11231).
H. H. and X. J. L. were supported by the Australian Research Council (ARC) (Grants Nos. FT140100003, FT130100815, DP140103231, and DP140100637).


\begin{thebibliography}{99}
\bibitem{BCS-BEC}          {D. M. Eagles, Phys. Rev. {\bf 186}, 456 (1969);
                            A. J. Leggett, in \emph{Modern trends in the theory of condensed matter}, Springer-Verlag, Berlin, 1980;
                            P. Nozieres and S. Schmitt-Rink, J. Low. Temp. Phys. {\bf 59}, 195 (1985);
                            C. A. R. Sa de Melo, M. Randeria, and J. R. Engelbrecht, Phys. Rev. Lett. {\bf 71}, 3202 (1993);
                            Q. Chen, J. Stajic, S. Tan, and K. Levin, Phys. Rep. {\bf 412}, 1 (2005).}
\bibitem{EXP}              {M. Greiner, C. A. Regal, and D. S. Jin, Nature {\bf 426}, 537 (2003);
                            S. Jochim, M. Bartenstein, A. Altmeyer, G. Hendl, S. Riedl, C. Chin, J. H. Denschlag, and R. Grimm,
                            Science {\bf 302}, 2101 (2003);
                            M. W. Zwierlein, J. R. Abo-Shaeer, A. Schirotzek, C. H. Schunck, and W. Ketterle, Nature {\bf 435}, 1047 (2005).}
\bibitem{Feshbach}         {T. Koehler, K. Goral, and P. S. Julienne, Rev. Mod. Phys. {\bf 78}, 1311 (2006);
                            C. Chin, R. Grimm, P. Julienne, and E. Tiesinga, Rev. Mod. Phys. {\bf 82}, 1225 (2010).}
\bibitem{book}             {\emph{Bose-Einstein Condensation in Dilute Gases}, C. J. Pethick and H. Smith, Cambridge University Press, 2002.}
\bibitem{QMC}              {J. Carlson,  S.-Y. Chang, V. R. Pandharipande, and K. E. Schmidt, Phys. Rev. Lett. {\bf 91}, 050401 (2003);
                            C. Lobo, A. Recati, S. Giorgini, and S. Stringari, Phys. Rev. Lett. {\bf 97}, 200403 (2006);
                            M. M. Forbes, S. Gandolfi, and A. Gezerlis,  Phys. Rev. Lett. {\bf 106}, 235303 (2011);
                            S. Giorgini, L. P. Pitaevskii, and S. Stringari, Rev. Mod. Phys. {\bf 80}, 1215 (2008).}
\bibitem{Expnarrow}        {E. L. Hazlett,  Y. Zhang, R. W. Stites, and K. M. O'Hara, Phys. Rev. Lett. {\bf 108}, 045304 (2012).}
\bibitem{atom-molecule01}  {E. Timmermans, P. Tommasini, M. Hussein, and A. Kerman, Phys. Rep. {\bf 315}, 199 (1999).}
\bibitem{atom-molecule02}  {M. Holland, S. J. J. M. F. Kokkelmans, M. L. Chiofalo, and R. Walser, Phys. Rev. Lett. {\bf 87}, 120406 (2001);
                            S. J. J. M. F. Kokkelmans,  J. N. Milstein, M. L. Chiofalo, R. Walser, and M. J. Holland,
                            Phys. Rev. {\bf A65}, 053617 (2002);
                            J. N. Milstein, S. J. J. M. F. Kokkelmans, and M. J. Holland, Phys. Rev. {\bf A66}, 043604 (2002).}
\bibitem{atom-molecule03}  {Y. Ohashi and A. Griffin, Phys. Rev. {\bf A67}, 063612(2003).}
\bibitem{atom-molecule04}  {V. Gurarie and L. Radzihovsky, Ann. Phys. (N. Y.) {\bf 322}, 2 (2007).}
\bibitem{barrier}          {S. De Palo, M. L. Chiofalo, M. J. Holland, and S. J. J. M. F. Kokkelmans, Phys. Lett. {\bf A327}, 490 (2004);
                            L. M. Jensen, H. M. Nilsen, and G. Watanabe, Phys. Rev. {\bf A74}, 043608 (2006).}
\bibitem{Bruun}            {G. M. Bruun, A. D. Jackson, and E. E. Kolomeitsev, Phys. Rev. {\bf A71}, 052713 (2005).}
\bibitem{Zhai}             {R. Qi and H. Zhai, Phys. Rev. {\bf A85}, 041603(R) (2012).}
\bibitem{two-band}         {H. Suhl, B. T. Matthias, and L. R. Walker, Phys. Rev. Lett. {\bf 3}, 552 (1959).}
\bibitem{two-band-iskin}   {M. Iskin and C. A. R. Sa de Melo, Phys. Rev. {\bf B72}, 024512 (2005); Phys. Rev. {\bf B74}, 144517 (2006).}
\bibitem{Hu}               {H. Hu,  X.-J. Liu, and P. D. Drummond, Europhys. Lett. {\bf 74}, 574 (2006); Nat. Phys. {\bf 3}, 469 (2007);
                            R. B. Diener, R. Sensarma, and M. Randeria, Phys. Rev. {\bf A77}, 023626 (2008).}
\bibitem{NSR-narrow}       {B. Marcelis and S. Kokkelmans, Phys. Rev. {\bf A74}, 023606 (2006);
                            T.-L. Ho, X. Cui, and W. Li, Phys. Rev. Lett. {\bf 108}, 250401 (2012).}
\bibitem{Zhu}              {G. Zhu and A. J. Leggett, Phys. Rev. {\bf A87}, 023627 (2013).}
\bibitem{He}               {L. He and P. Zhuang, Phys. Rev. {\bf B79}, 024511 (2009).}
\end{thebibliography}
\end{document}